\documentclass[12pt]{article}
\pdfoutput=1
\usepackage{subfigure}
\usepackage{amssymb,amsmath}
\usepackage{graphicx}
\usepackage{pbox}
\usepackage{color}
\usepackage{cancel}
\usepackage[colorlinks=true
,urlcolor=blue
,citecolor=blue
,linkcolor=blue
,pagecolor=blue
,linktocpage=true
,pdfproducer=medialab
]{hyperref}
\usepackage[a4paper,width=15.2cm]{geometry}

\usepackage[section]{placeins}
\usepackage{ulem}
\usepackage{cite}

\usepackage{enumitem} 

\makeatletter \renewcommand{\@dotsep}{10000} \makeatother
\def\be{\begin{equation}}
\def\ee{\end{equation}}
\def\bea{\begin{eqnarray}}
\def\eea{\end{eqnarray}}
\def\bi{\begin{itemize}}
\def\ei{\end{itemize}}

\newcommand\prd[3]{{\it Phys.\ Rev.\ }{\bf D #1} (#2) #3}
\newcommand\prl[3]{{\it Phys.\ Rev.\ Lett.\ }{\bf #1} (#2) #3}

\newcommand{\beq}{\begin{equation}}
\newcommand{\eeq}{\end{equation}}

\newcommand{\Rmnum}[1]{\expandafter\@slowromancap\romannumeral #1@}

\begin{document}
%Remove date before submitting to arXiv
\date{\today}

\begin{center}
{\Large\bf 
Neutralino Dark Matter and Other LHC Predictions from Quasi Yukawa Unification 
} \vspace{1cm}
\end{center}

\begin{center}
{\Large Qaisar Shafi$^a$\footnote{Email: shafi@bartol.udel.edu}, \c{S}\"{u}kr\"{u} Hanif Tany\i ld\i z\i$^b$\footnote{Email: hanif@theor.jinr.ru} and Cem Salih \"{U}n$^c$\footnote{Email: cemsalihun@uludag.edu.tr}}

\vspace{0.75cm}

\hspace{-3.0cm}{\it $^a$ \small{Bartol Research Institute, Department of Physics and Astronomy,\\
University of Delaware, Newark, DE 19716, USA}} 

\vspace{0.2cm}

{\it $^b$ \small{Bogoliubov Laboratory of Theoretical Physics, Joint Institute for Nuclear Research, \\ \hspace{-1.8cm}141980, Dubna, Moscow Region, Russia}}

\vspace{0.2cm}

\hspace{-2.5cm}{\it $^c$ \small{Department of Physics, Uluda\u{g} University, Bursa, Turkey,  TR16059}}

%
%{\Large
%Shabbar Raza$^{a,b,}$\footnote{
%Email: shabbar@itp.ac.cn},
%
%Qaisar Shafi $^{b,}$\footnote{
%Email: shafi@bartol.udel.edu.
%}
%
%and
%
%Cem Salih Un $^{b,c,}$\footnote{
%Email: cemsalihun@bartol.udel.edu.}
%}

%\vspace{0.75cm}
%
%{\it $^a$
%State Key Laboratory of Theoretical Physics and Kavli Institute for Theoretical Physics China (KITPC),
%Institute of Theoretical Physics, Chinese Academy of Sciences, Beijing 100190, P. R. China
%}\\
%
%{\it  $^b$
%Bartol Research Institute, Department of Physics and Astronomy,\\
%University of Delaware, Newark, DE 19716, USA %\\ \vspace{2mm}
%}
%

%{\it  $^c$
%Uludag University, Faculty of Arts \& Science, \\
%G\"or\"ukle Campus, Nil\"ufer Bursa 16059, Turkey}
%

\vspace{1.5cm}
\section*{Abstract}
\end{center}

We explore the dark matter and LHC implications of $t-b-\tau$ quasi Yukawa unification in the framework of supersymmetric models based on the gauge symmetry $G=SU(4)_{c}\times SU(2)_{L}\times SU(2)_{R}$. The deviation from exact Yukawa unification is quantified by a dimensionless parameter $C$ ($|C| \lesssim 0.2$), such that the Yukawa couplings at $M_{{\rm GUT}}$ are related by $y_{t}:y_{b}:y_{\tau}=|1+C| : |1-C| : |1+3C|$. In contrast to earlier studies which focused on universal gaugino masses, we consider non-universal gaugino masses at $M_{{\rm GUT}}$ that are compatible with the gauge symmetry $G$. We perform two independent scans of the fundamental parameter space, one of which employs ISAJET, while the other uses SoftSusy interfaced with SuperIso. These scans reveal qualitatively similar allowed regions in the parameter space, and yield a variety of neutralino dark matter scenarios consistent with the observations. These include stau and chargino coannihilation scenarios, the $A-$resonance scenario, as well as Higgsino dark matter solution which is more readily probed by direct detection searches. The gluino mass is found to be $\lesssim 4.2$ TeV, the stop mass is $\gtrsim 2$ TeV, while the first two family squarks and sleptons are of order $4-5$ TeV and $3$ TeV respectively.

%\textbf{We consider quasi Yukawa unification in the framework of Pati-Salam model in which the Yukawa couplings receive contributions from the Higgs fields in (15,1,3) multiplet of $SU(4)_{c}\times SU(2)_{L}\times SU(2)_{R}$. These contributions} result in breaking \textbf{exact} Yukawa unification (YU). We restrict such contributions to $C\lesssim 0.2$ referred to the Quasi Yukawa Unification (QYU), where C \textbf{stands for} the deviation from the exact Yukawa Unification, and discuss the allowed parameter space, mass spectrum and the dark matter implications of the QYU consistent with the experimental constraints. We perform two different scans one of which employs ISAJET, while the other uses SoftSusy interfaced with SuperIso. These scans reveal similar regions in the fundamental parameter space as well as the phenomenological implications within reach of experiments at the Large Hadron Collider (LHC). We find the gluino mass $\sim 1.2-4.2$ TeV and the stop mass $\gtrsim 2$ TeV. Stau and chargino coannihilation channels and A-funnel solutions can be realized to reduce the relic abundance of neutralino LSP to the desired range to be a consistent dark matter candidate. Also we find the higgsino dark matter solutions that are testable by the current direct detection experiments.

\newpage

\section{Introduction}
\label{sec:intro}

In an earlier paper \cite{Dar:2011sj}, hereafter referred to as I, we have explored the LHC implications of imposing $t-b-\tau$ Quasi-Yukawa Unification (QYU) at the grand unification scale ($M_{{\rm GUT}}\sim 2\times 10^{16}$ GeV). This modified approach to the third family ($t-b-\tau$) YU \cite{big-422} can be motivated by the desire to construct realistic supersymmetric models of grand unified theories (GUTs) which also incorporate realistic masses and mixings observed in the matter sector. For instance, the desired quarks and charged lepton masses for the second family fermion can be incorporated, following \cite{Gomez:2002tj}, by including Higgs fields in the (15,1,3) representation of $G=SU(4)_{c}\times SU(2)_{L}\times SU(2)_{R}$ (4-2-2) \cite{pati-salam}, which develop a non-zero GUT scale vacuum expectation values (VEVs). The third family Yukawa couplings receive, in this case, sizable new contributions, and the deviations from exact YU can be stated as follows \cite{Dar:2011sj}:

\begin{equation}
y_{t}:y_{b}:y_{\tau}=\mid 1+C \mid : \mid 1-C \mid : \mid 1+3C \mid 
\label{QYU}
\end{equation}
where $C$ measures the deviation from the exact YU. Restricting the deviation to $C \lesssim 0.2$ we refer Eq.(\ref{QYU}) to QYU condition. 

%After discovery of the Higgs boson of mass about 125-126 GeV independently reported by the ATLAS \cite{Aad:2012tfa} and CMS \cite{Chatrchyan:2012ufa} experiments, it has brought strong hints for physics beyond the Standard Model (SM). Supersymmetry (SUSY) is one of the forefront candidates, and minimal supersymmetric extension of the SM (MSSM) is consistent with observations on the Higgs boson, since MSSM predicts $m_{h} \lesssim 135$ GeV, and the lightest CP-even Higgs boson, one of the five physical Higgs boson states in MSSM, exhibits very similar properties at the decoupling limit \cite{Haber:1993pv}. Besides, the MSSM has plausible candidates for the dark matter (DM) when R-parity is imposed that requires the lightest supersymmetric particle (LSP) to be stable. In addition, it unifies all three gauge couplings of the SM gauge groups, and since it ensures cancellation of infinitive loop contributions to the Higgs mass, one can consider the high scale theories and explore their low scale implications through renormalization group equations (RGEs) within the MSSM. Among such theories $SO(10)$ grand unified theories (GUTs) \cite{big-422} and Pati-Salam model \cite{pati-salam} ($SU(4)_{c}\times SU(2)_{L}\times SU(2)_{R}$, hereafter $4-2-2$ in short) form a special class by predicting Yukawa unification (YU) at the GUT-scale ($M_{{\rm GUT}}$). 

The 4-2-2 model has many salient features distinguishing it from other high scale theories \cite{Lazarides:1980tg}. The discrete left-right (LR) symmetry reduces the number of gauge couplings from three to two with $g_{L}=g_{R}$. It also requires gaugino masses of $SU(2)_{L}$ and $SU(2)_{R}$ to be equal at $M_{{\rm GUT}}$. The matter fields of each family belong to $\psi(4,2,1)$ and $\psi_{c}(\bar{4},1,2)$. The LR symmetry requires the existence of right handed neutrino. 

In this paper we reconsider QYU in the framework of 4-2-2 defined above, taking into account the fact that the MSSM gaugino masses $M_{1,2,3}$ at $M_{{\rm GUT}}$ can be non-universal. In particular, we assume the following asymptotic relation \cite{Gogoladze:2009ug}

 \begin{equation}
M_{1} = \frac{3}{5}M_{2} + \frac{2}{5}M_{3} ~~,
\label{gaugino422}
\end{equation}
which follows from the assumption of left-right symmetry at $M_{{\rm GUT}}$ and the fact that $U(1)_{Y}$ derived from 4-2-2 is given as follows;

\begin{equation}
Y = \sqrt{\frac{3}{5}}I_{3R} + \sqrt{\frac{2}{5}}(B-L).
\label{hyper422}
\end{equation}
Here $M_{1}$, $M_{2}$ and $M_{3}$ are the asymptotic gaugino masses for $U(1)_{Y}$, $SU(2)_{L}$ and $SU(3)_{c}$, and $I_{3R}$ and $(B-L)$ are diagonal generators of $SU(2)_{R}$ and $SU(4)_{c}$ respectively.

The setup of our paper is as follows. In Section \ref{sec:scan} we briefly explain our scanning procedure and list the experimental constraints that we impose on the data obtained from our scans. In Section \ref{sec:spec} we show the fundamental parameter space that is allowed by the experimental constraints and QYU. Section \ref{sec:coan} provides the implications for dark matter sector such as coannihilation channels and the resonance solution. Section \ref{sec:higgsinoDM} considers a Higgsino-like  LSP and emphasize the implications for direct detection experiments. Section \ref{sec:isasoft} compares the ISAJET and SoftSusy, and the small variations in their phenomenology. We also present the benchmark points obtained from the different scans to exemplify our results. Finally, we conclude our study in Section \ref{sec:conclusion}. 

\section{Scanning Procedure and Experimental \\ Constraints}
\label{sec:scan}

In our scan, we employ ISAJET 7.84 \cite{Paige:2003mg}, SoftSusy-3.4.1 \cite{Allanach:2001kg} and SuperIso Relic v3.3 \cite{Mahmoudi:2008tp} to calculate the low scale observables. The gauge and Yukawa couplings are first estimated at the low scale. ISAJET evolves the gauge couplings and the Yukawa couplings of the third family up to $M_{{\rm GUT}}$, while SoftSusy performs the calculations in the three family approximation in evolution of Yukawa couplings. We do not strictly enforce the gauge unification condition $g_{1}=g_{2}=g_{3}$, since a few percent deviation from the gauge coupling unification can be generated by unknown GUT-scale threshold corrections \cite{Hisano:1992jj}. Hence, $M_{{\rm GUT}}$ is calculated to be the scale where $g_{1}=g_{2}$ and $g_{3}$ deviates a few percent. After $M_{{\rm GUT}}$ is determined, the soft supersymmetry breaking (SSB) parameters determined with the boundary conditions defined at $M_{{\rm GUT}}$ are evolved together with the gauge and Yukawa couplings from $M_{{\rm GUT}}$ to the weak scale $M_{{\rm Z}}$.

The SUSY threshold corrections to the Yukawa couplings \cite{Pierce:1996zz} are taken into account at the common scale $M_{{\rm SUSY}} = \sqrt{m_{\tilde{t}_{L}}m_{\tilde{t}_{R}}}$ in ISAJET, while SoftSusy evaluates them at the electroweak scale. The entire parameter set is iteratively run between $M_{{Z}}$ and $M_{{\rm GUT}}$ using full 2-loop RGEs, and the SSB parameters are extracted from RGEs at the appropriate scales $m_{i} = m_{i}(m_{i})$. 

We have performed random scans over the following parameter space:

\begin{eqnarray}
\label{paramspace}
0 \leq m_{16} \leq 10000 ~{\rm GeV} \nonumber \\
0 \leq M_{2} \leq 2000 ~{\rm GeV} \nonumber \\
0 \leq M_{3} \leq 2000 ~{\rm GeV} \nonumber \\
-3 \leq A_{0}/m_{16} \leq 3  \\
40 \leq \tan \beta \leq 60 \nonumber \\
0 \leq m_{10} \leq 10000 ~{\rm GeV} \nonumber \\
\mu > 0, \hspace{0.5cm} m_{t} = 173.3 ~{\rm GeV}, \nonumber
\end{eqnarray}
taking Eq.\ref{gaugino422} into account. We use the $SO(10)$ notation in which $m_{16}$ is the universal SSB mass term for the particles, and $M_{1}$, $M_{2}$, $M_{3}$ are SSB mass terms for the gauginos of $U(1)_{Y}$, $SU(2)_{L}$ and $SU(3)_{c}$ respectively, $A_{0}$ is the universal SSB term for trilinear scalar interactions, $\tan \beta$ is the ratio of VEVs of the MSSM higgs doublets, $\mu$ is coefficient of the bilinear Higgs mixing term, and $m_{t}$ is the top quark mass. Note that we set the top quark mass to 173.3 GeV \cite{ATLAS:2014wva, Gogoladze:2014hca}, and our results are not too sensitive to a $1\sigma - 2\sigma$ variation in $m_{t}$ \cite{bartol2}.

We employ the Metropolis-Hasting algorithm as described in \cite{Belanger:2009ti}, and require all collected points to satisfy radiative electroweak symmetry breaking (REWSB) with LSP neutralino. The REWSB gives a crucial theoretical constraint on the parameter space \cite{Ibanez:1982fr}. After collecting data, we impose constraints from the mass bounds \cite{Nakamura:2010zzi}, rare decays of B-meson such as $B_{s}\rightarrow \mu^{+}\mu^{-}$ \cite{Aaij:2012nna}, $b\rightarrow s \gamma$ \cite{Amhis:2012bh}, and $B_{u}\rightarrow \tau \nu_{\tau}$ \cite{Asner:2010qj}. After obtaining the region allowed by the LHC constraints, we also apply the WMAP bound \cite{WMAP9} on the relic abundance of LSP neutralino. ISAJET interfaces with IsaTools \cite{bsg,bmm} for B-physics and relic density observables, while we interface SoftSusy with SuperIso Relic in order to calculate these observables. The experimental constraints imposed in our data can be summarized as follows:

\begin{eqnarray}
\label{constraints}
m_{h}=(123-127)~{\rm GeV} \nonumber \\
m_{\tilde{g}} \leq 1~{\rm TeV} \nonumber \\
0.8\times10^{-9} \leq {\rm BR}(B_s \rightarrow \mu^+ \mu^-) \leq 6.2\times 10^{-9}~(2\sigma) \nonumber \\
2.99\times 10^{-4} \leq {\rm BR}(b \rightarrow s \gamma) \leq 3.87\times 10^{-4}~(2\sigma) \\
0.15 \leq \frac{
 {\rm BR}(B_u\rightarrow\tau \nu_{\tau})_{\rm MSSM}}
 {{\rm BR}(B_u\rightarrow \tau \nu_{\tau})_{\rm SM}} \leq 2.41 ~(3\sigma) \nonumber \\
 0.0913 \leq \Omega_{{\rm CDM}}h^{2} \leq 0.1363~(5\sigma). \nonumber
\end{eqnarray}

We emphasize here the mass bounds on the Higgs boson \cite{Aad:2012tfa,Chatrchyan:2012ufa} and gluino \cite{TheATLAScollaboration:2013fha}. We allow a few percent deviation from the observed mass of the Higgs boson, since there exist about a 2 GeV error in estimation of its mass arising due to theoretical uncertainties in the calculation of the minimum of the scalar potential, and experimental uncertainties in $m_{t}$ and $\alpha_{s}$ \cite{Degrassi:2002fi}. Besides these constraints, we require our solutions to do  no worse than the SM in comparing predictions for the muon anomalous magnetic moment. Note that we relax the WMAP bound to $0.0913 \leq \Omega_{{\rm CDM}}h^{2} \leq 1$ on the solutions obtained from SuperIso Relic in order to take into account uncertainties in the calculation of the relic abundance of LSP neutralino.

\section{Fundamental Parameter Space of Quasi Yukawa Unification and Sparticle Mass Spectrum }
\label{sec:spec}

\begin{figure}[t!]
\begin{equation*}\hspace{-0.8cm}
\begin{array}{cc}
{\rm ISAJET} & ~~~~~~~~~~~{\rm SoftSusy~ and~ SuperIso Relic} \\ \\
\includegraphics[scale=1]{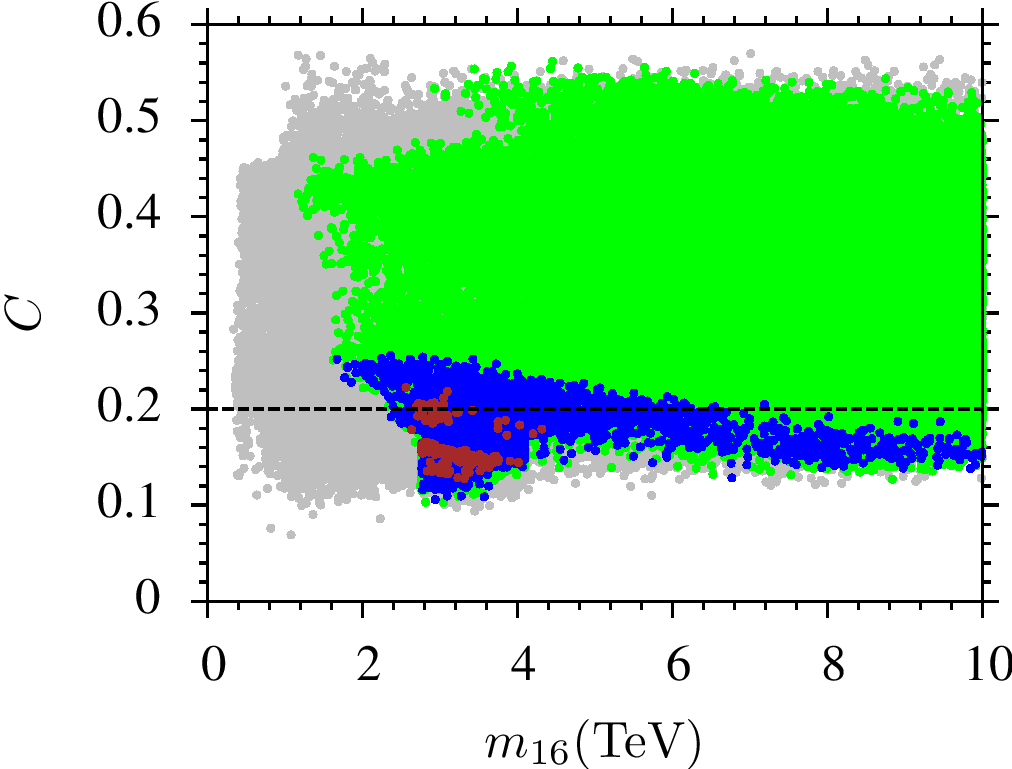} &
\includegraphics[scale=1]{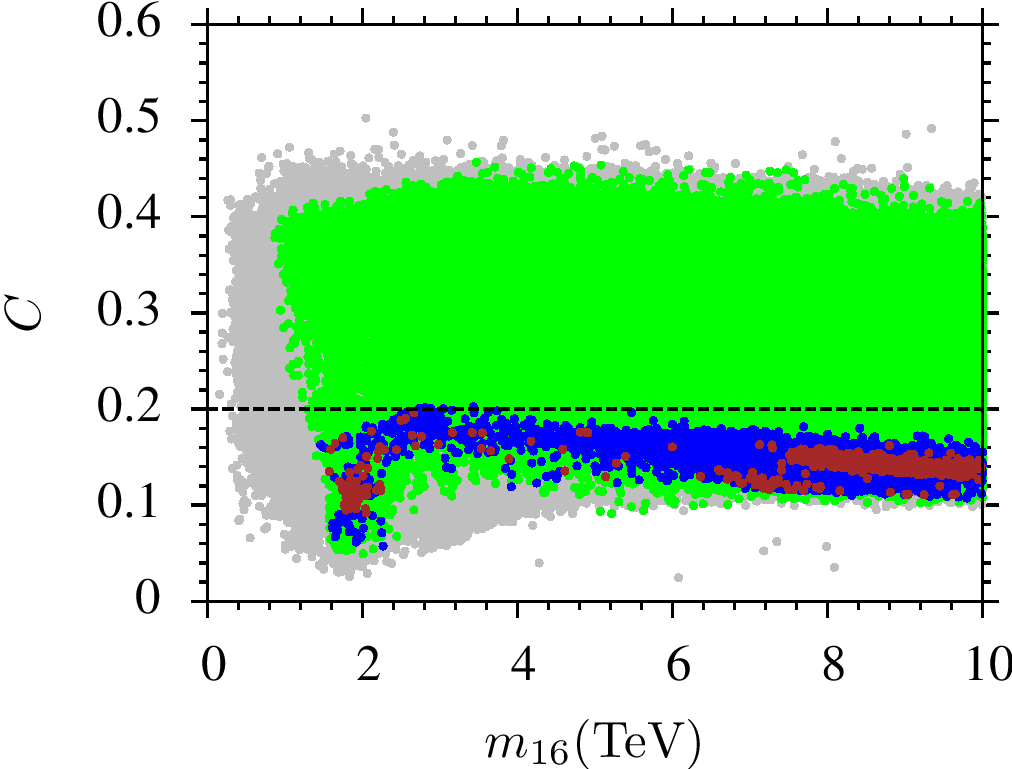} \\
\includegraphics[scale=1]{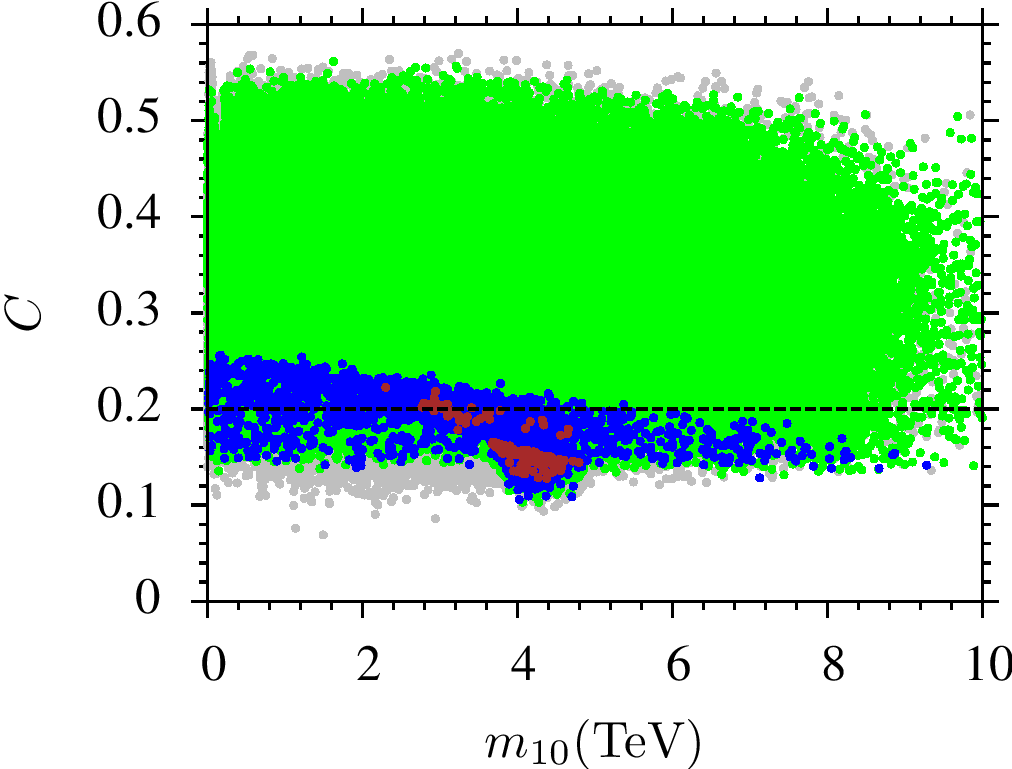} &
\includegraphics[scale=1]{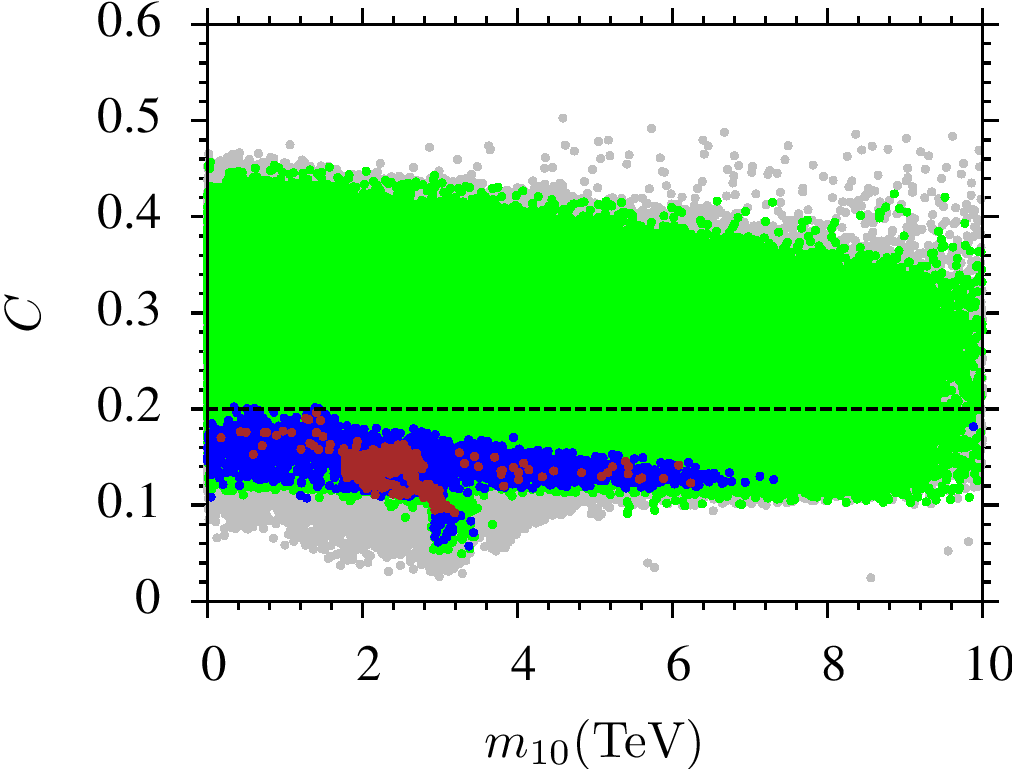} 
\end{array}
\end{equation*}
\caption{Plots in $C-m_{16}$ and $C-m_{10}$ planes. The left panel shows the results obtained by using ISAJET, while the right panels display the results from SoftSusy. All points are compatible with REWSB and LSP neutralino. Green points satisfy the mass bounds on the sparticles and the constraints from rare B-decays. Blue points form a subset of green and they are compatible with the QYU condition. Points in red form a subset of blue and satisfy the constraint on relic abundance of LSP neutralino. They are consistent with the WMAP bound within $5\sigma$ uncertainty in ISAJET plots, while $0.0913 \leq \Omega h^{2} \leq 1$ for those obtained from SoftSusy and SuperIso Relic.}
\label{C_mass}
\end{figure}

\begin{figure}[t!]
\begin{equation*}\hspace{-0.8cm}
\begin{array}{cc}
{\rm ISAJET} & ~~~~~~~~~~~{\rm SoftSusy~ and~ SuperIso Relic} \\  \\
\includegraphics[scale=1.2]{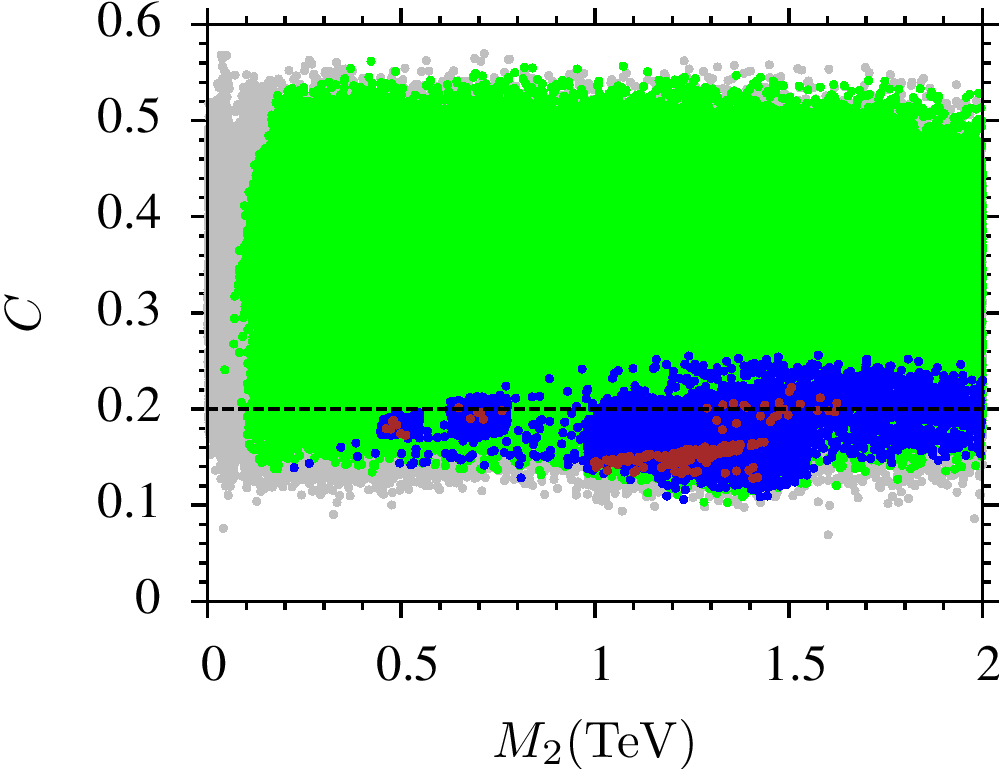} &
\includegraphics[scale=1.2]{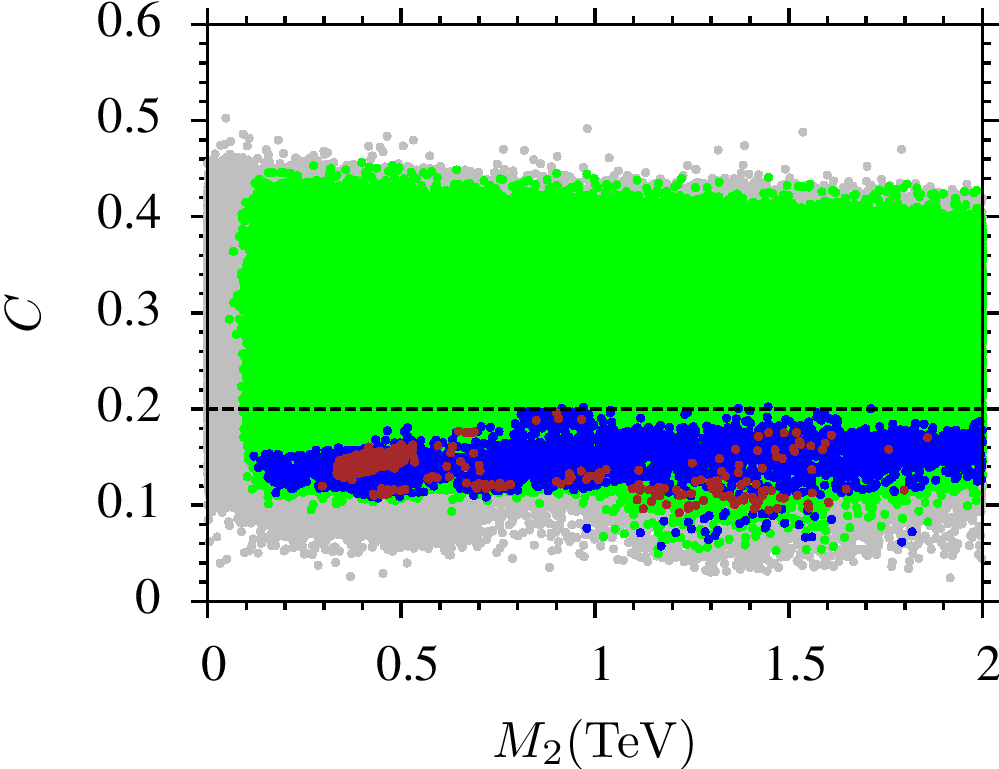} \\
\includegraphics[scale=1.2]{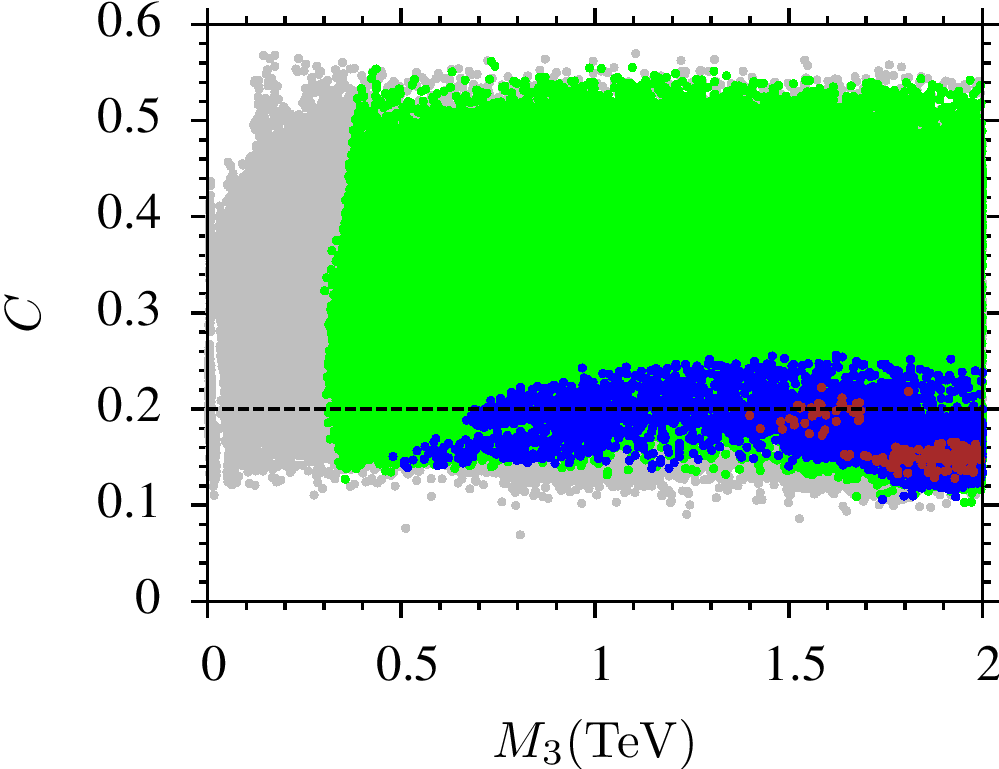} &
\includegraphics[scale=1.2]{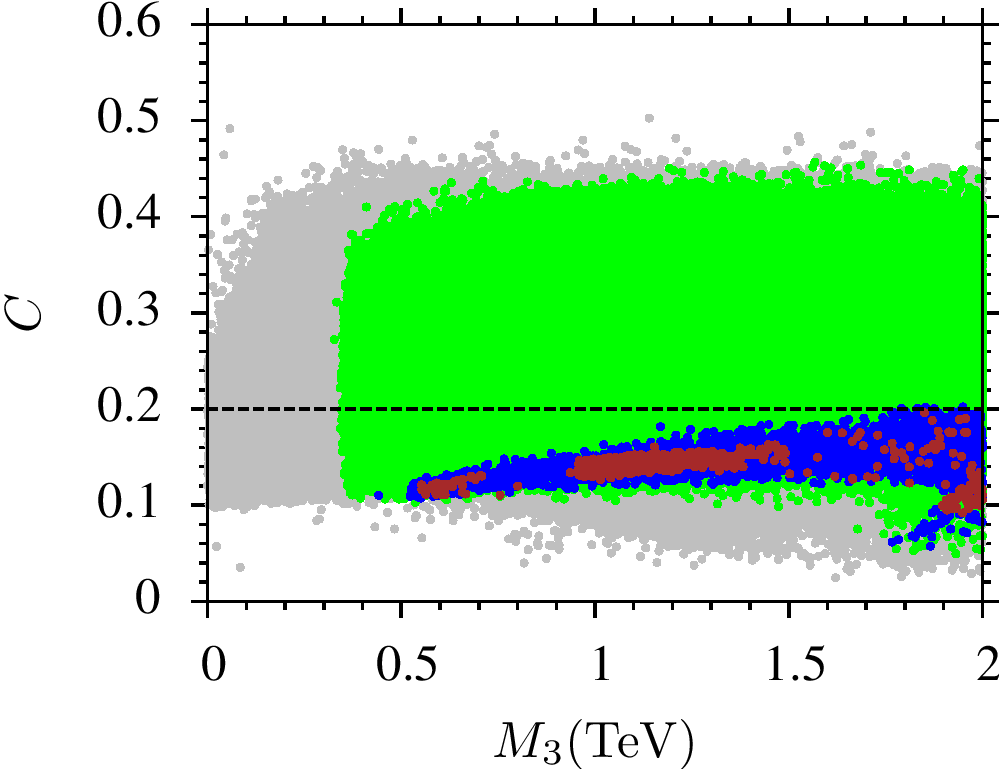} 
\end{array}
\end{equation*}
\caption{Plots in $C-M_{2}$ and $C-M_{3}$ planes. Color coding is the same as in Fig.\ref{C_mass}.}
\label{C_mass2}
\end{figure}

\begin{figure}[t!]
\begin{equation*}\hspace{-0.8cm}
\begin{array}{cc}
{\rm ISAJET} & ~~~~~~~~~~~{\rm SoftSusy~ and~ SuperIso Relic} \\ \\
\includegraphics[scale=1.2]{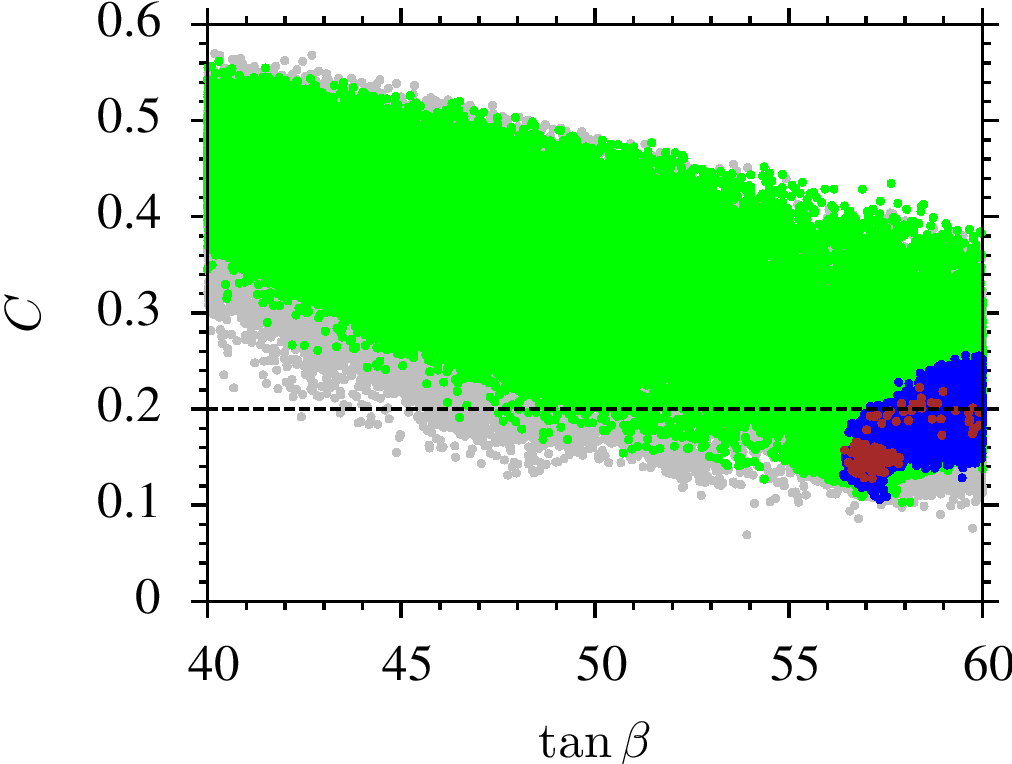} &
\includegraphics[scale=1.2]{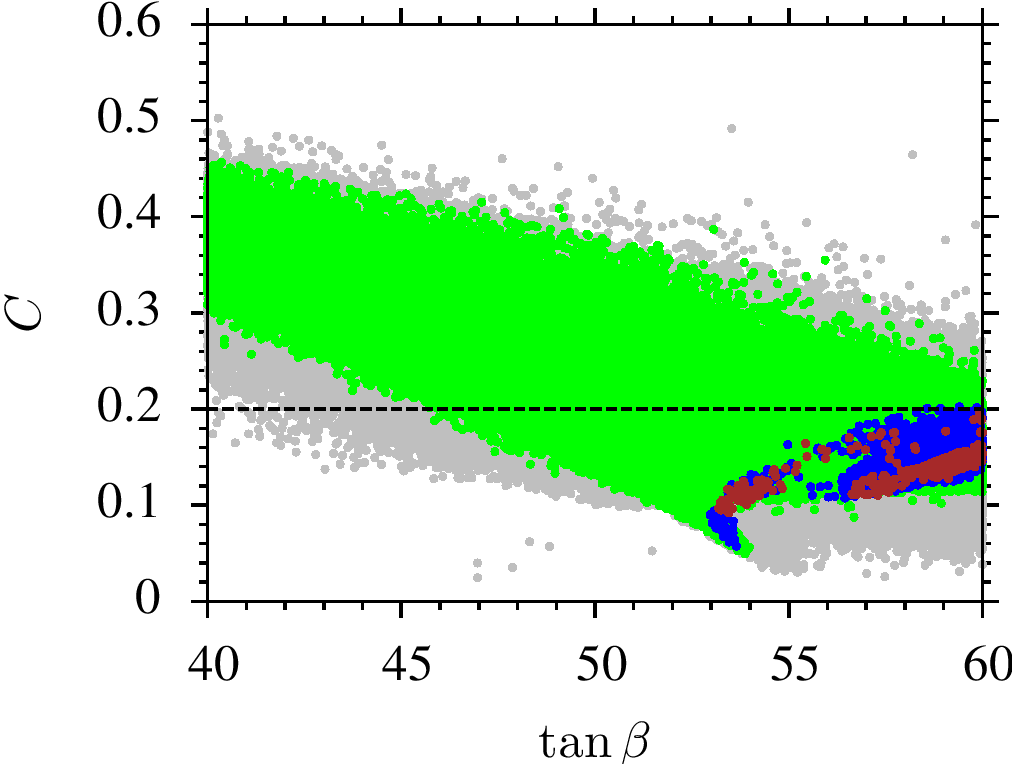} \\
\includegraphics[scale=1.2]{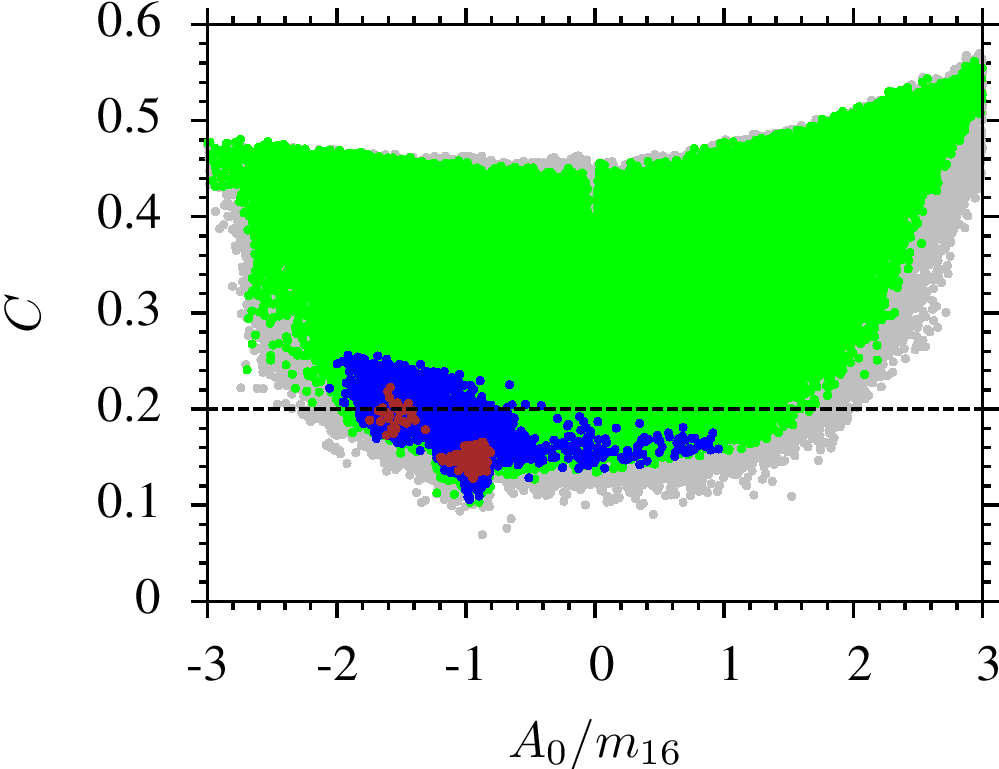} &
\includegraphics[scale=1.2]{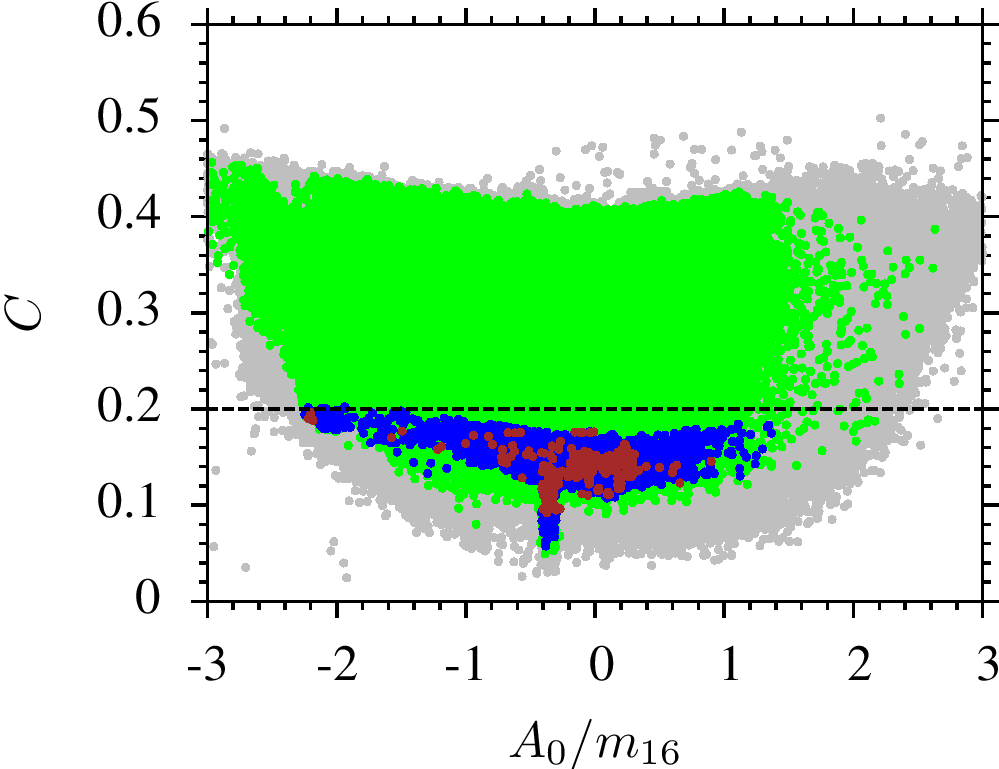}
\end{array}
\label{C_tA}
\end{equation*}
\caption{Plots in $C-\tan\beta$ and $C-A_{0}/m_{16}$ planes. Color coding is the same as in Fig.\ref{C_mass}.}
\end{figure}

%\begin{figure}[h!]
%\begin{equation*}\hspace{-0.8cm}
%\begin{array}{cc}
%{\rm ISAJET} & {\rm SoftSusy~ and~ SuperIso Relic} \\ 
%\includegraphics[scale=1.2]{422_NUHM1_quasi_C_gluino.png} &
%\includegraphics[scale=1.2]{422_NUHM1_quasi_soft_C_gluino.png} \\
%\includegraphics[scale=1.2]{422_NUHM1_quasi_C_mh0.png} &
%\includegraphics[scale=1.2]{422_NUHM1_quasi_soft_C_mh0.png}
%\end{array}
%\label{C_spec1}
%\end{equation*}
%\end{figure}

\begin{figure}[t!]
\begin{equation*}\hspace{-0.8cm}
\begin{array}{cc}
{\rm ISAJET} & ~~~~~~~~~~~{\rm SoftSusy~ and~ SuperIso Relic} \\ \\
\includegraphics[scale=1.2]{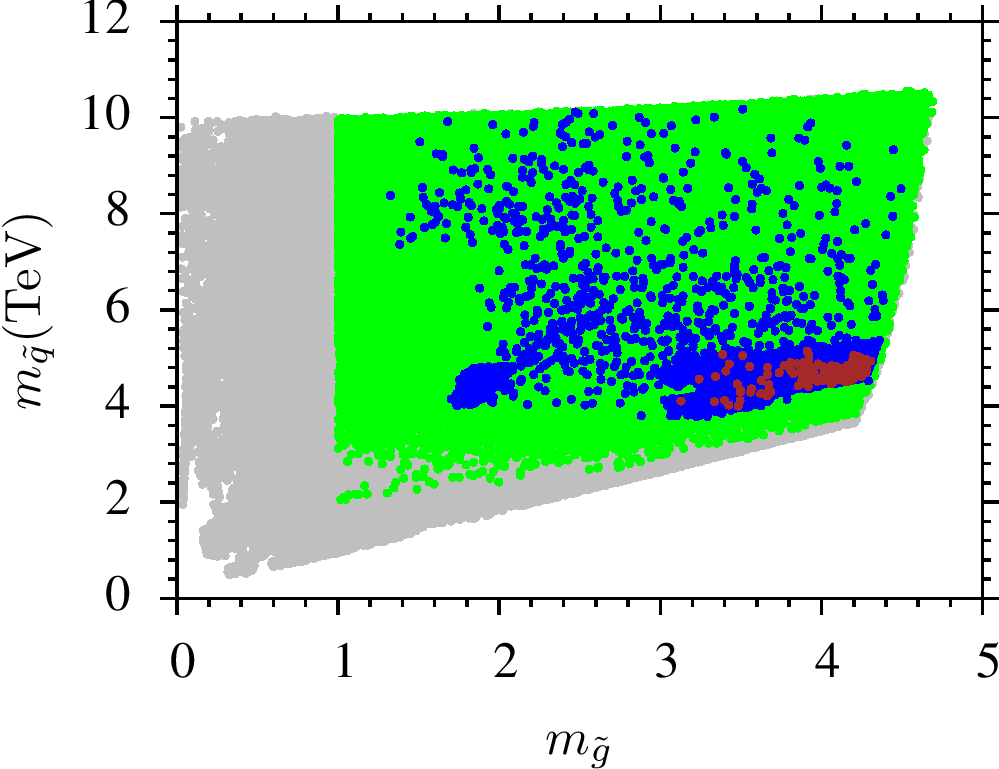} &
\includegraphics[scale=1.2]{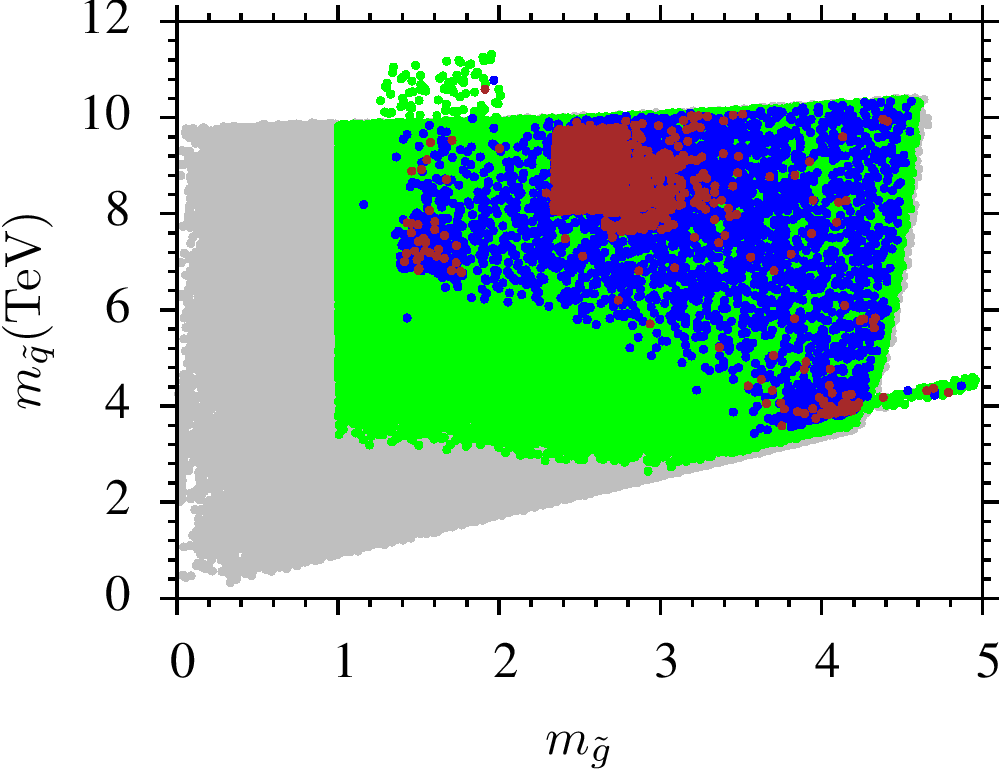} \\
\includegraphics[scale=1.2]{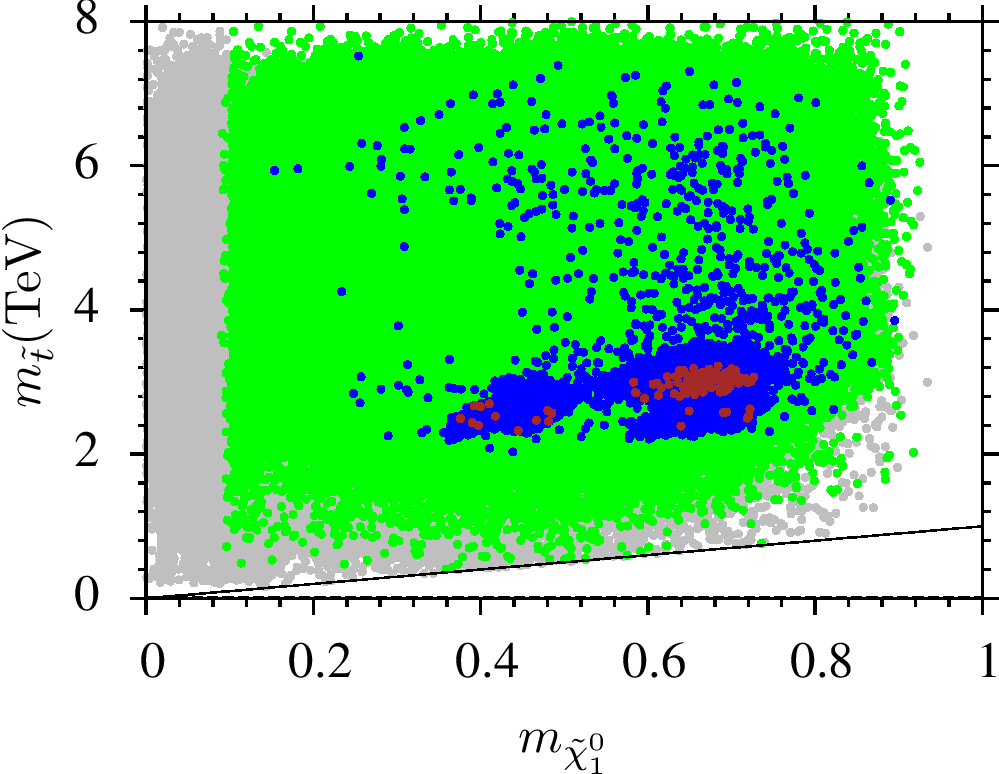} &
\includegraphics[scale=1.2]{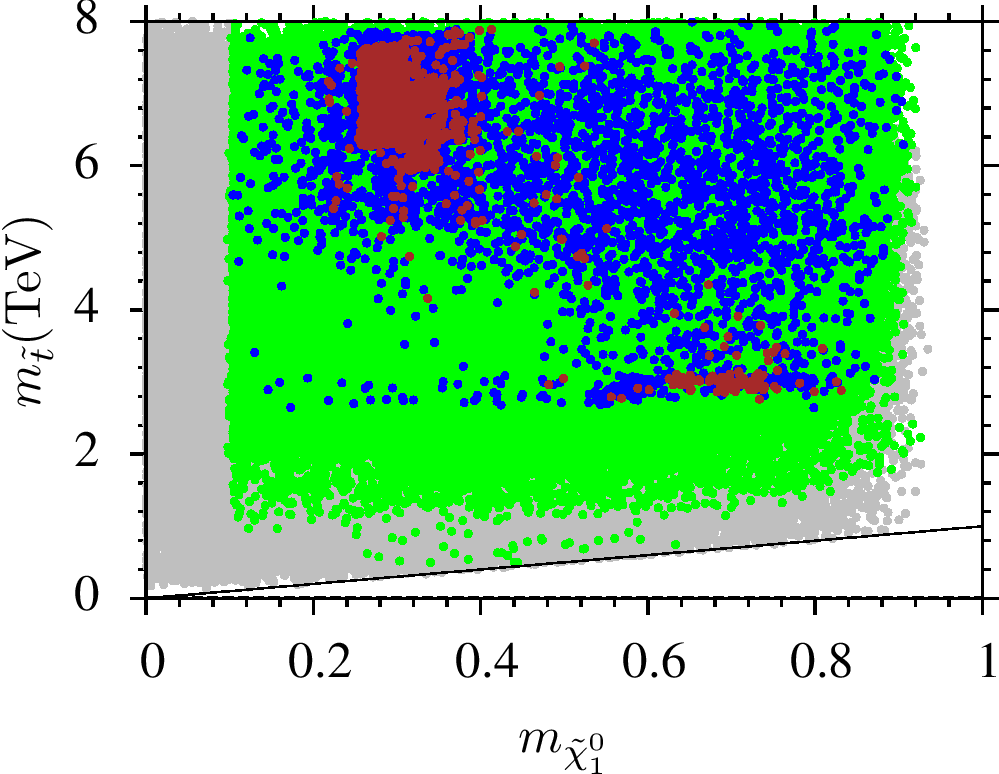}
\end{array}
\end{equation*}
\caption{Plots in $C-m_{\tilde{g}}$ and $C-m_{\tilde{t}}$ planes. Color coding is the same as in Fig.\ref{C_mass}. The blue points satisfy $C\leq 0.2$ as well as QYU condition.}
\label{C_spec}
\end{figure}

In this section, we highlight the allowed regions in the fundamental parameter space of 4-2-2 given in Eq.(\ref{paramspace}) and present the results for the supersymmetric particles and Higgs boson mass spectrum that we obtain from the scans using ISAJET and SoftSusy which is interfaced with SuperIso Relic. Fig.\ref{C_mass} shows the plots in $C-m_{16}$ and $C-m_{10}$ planes. The left panel shows the results obtained by using ISAJET, while the right panels displays the results from SoftSusy. All points are compatible with REWSB and LSP neutralino. Green points satisfy the mass bounds on the sparticles and the constraints from rare B-decays. Blue points form a subset of green and they are compatible with the QYU condition. Points in red are a subset of blue, and satisfy the constraint on relic abundance of LSP neutralino. They are consistent with the WMAP bound within $5\sigma$ uncertainty in ISAJET plots, while $0.0913 \leq \Omega h^{2} \leq 1$ for those obtained from SoftSusy and SuperIso Relic. As seen from the $C-m_{16}$ panels, QYU (blue) requires $m_{16} \gtrsim 2$ TeV, while $m_{10}$ is only loosely constrained.

Similarly Fig.\ref{C_mass2} displays the plots in $C-M_{2}$ and $C-M_{3}$ planes. Color coding is the same as in Fig.\ref{C_mass}. We can see from the $C-M_{2}$ panel that $M_{2}$ can be as low as 300 GeV. Such light $M_{2}$ solutions yield bino-wino mixing at the low scale which plays a role in reducing the relic abundance of LSP neutralino to the desired range. The $C-M_{3}$ plane shows that $M_{3} \gtrsim 500$ GeV is compatible with QYU which leads to a heavy gluino ($m_{\tilde{g}} \gtrsim 1.5$ TeV) at low scale. 

The results for the remaining parameters are shown in Fig.\ref{C_tA} with plots in $C-\tan\beta$ and $C-A_{0}/m_{16}$ planes. Color coding is the same as in Fig.\ref{C_mass}. The plots in the $C-\tan\beta$ planes shows that QYU requires rather high $\tan\beta$ values. The top left panel shows that $\tan\beta \gtrsim 56$ is compatible with QYU, while it is possible to find QYU solutions with SoftSusy for $\tan\beta \gtrsim 53$. The $C-A_{0}/m_{16}$ panels from both ISAJET and SoftSusy show that $A_{0}/m_{16}$ can lie in the range ($-2,1$). 

We present the results for the mass spectrum of the colored particles in Fig.\ref{C_spec} in $C-m_{\tilde{g}}$ and $C-m_{\tilde{t}}$ planes. Color coding is the same as in Fig.\ref{C_mass}. In addition, the blue points satisfy $C\leq 0.2$ as well as QYU condition. The gluino mass compatible with QYU and $C\leq 0.2$ is found to be $m_{\tilde{g}} \gtrsim 1.5$ TeV as stated above, and it can be tested in future experiments at the Large Hadron Collider (LHC). Similarly, the stop quarks satisfy $m_{\tilde{t}} \gtrsim 2$ TeV. ISAJET and SoftSusy are in good agreement regarding results of the mass spectrum.

\section{LSP Neutralino and Coannihilation Scenarios}
\label{sec:coan}

In the previous section we have focused on the fundamental parameter space and the mass spectrum of the colored particles. Since we accept only those solutions which lead to LSP neutralino, it is worth investigating the implications of 4-2-2 on the dark matter observables. Indeed, if the LSP neutralino is mostly a bino, its relic abundance is usually so high that it cannot be consistent with the WMAP observation. However, one can identify various coannihilation channels that reduce the relic abundance of neutralino to the desired ranges. 4-2-2 has some rich phenomenological implications and allows various coannihilation channel scenarios at the low scale \cite{Gogoladze:2009bn}, since it allows asymptotically different masses for the gauginos as given in Eq.(\ref{gaugino422}). On the other hand, if one imposes $t-b-\tau$ YU at $M_{{\rm GUT}}$ with $\mu > 0$, only the gluino-neutralino coannihilation channel can survive \cite{Gogoladze:2009ug}. Relaxing this to $b-\tau$ YU opens up, in addition, the stop-neutralino channel \cite{Shabbar}. In this section, we consider the phenomenological implications of QYU in 4-2-2 regarding the dark matter and the structure of LSP neutralino. Besides the bino-like LSP neutralino, it is possible to find solutions with bino-wino mixture, bino-higgsino mixture, or mostly higgsino LSP neutralino which leads to different phenomenology.

\begin{figure}[h!]
\begin{equation*}\hspace{-0.8cm}
\begin{array}{cc}
{\rm ISAJET} & ~~~~~~~~~~~{\rm SoftSusy~ and~ SuperIso Relic} \\ \\
\includegraphics[scale=1.2]{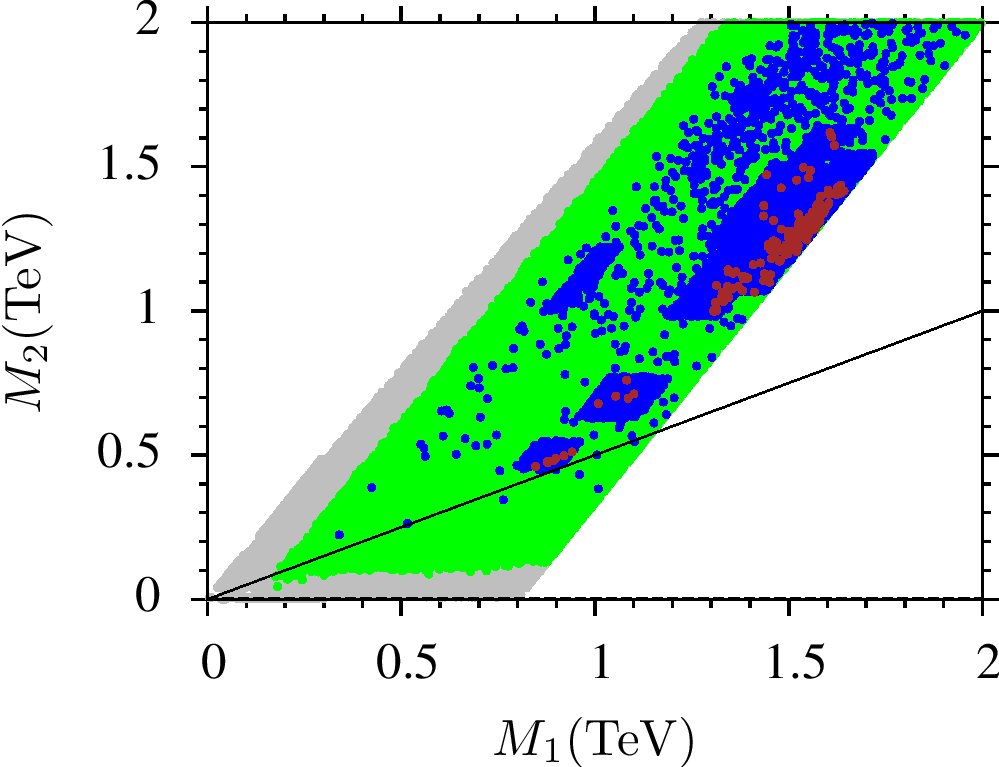} &
\includegraphics[scale=1.2]{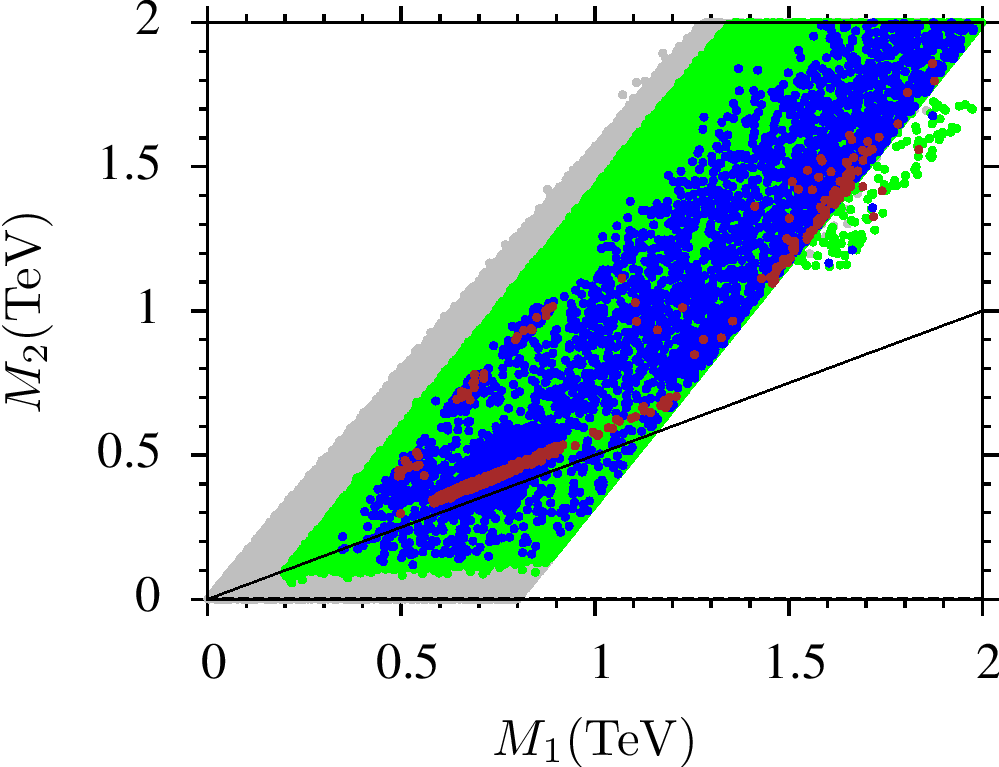} \\
\includegraphics[scale=1.2]{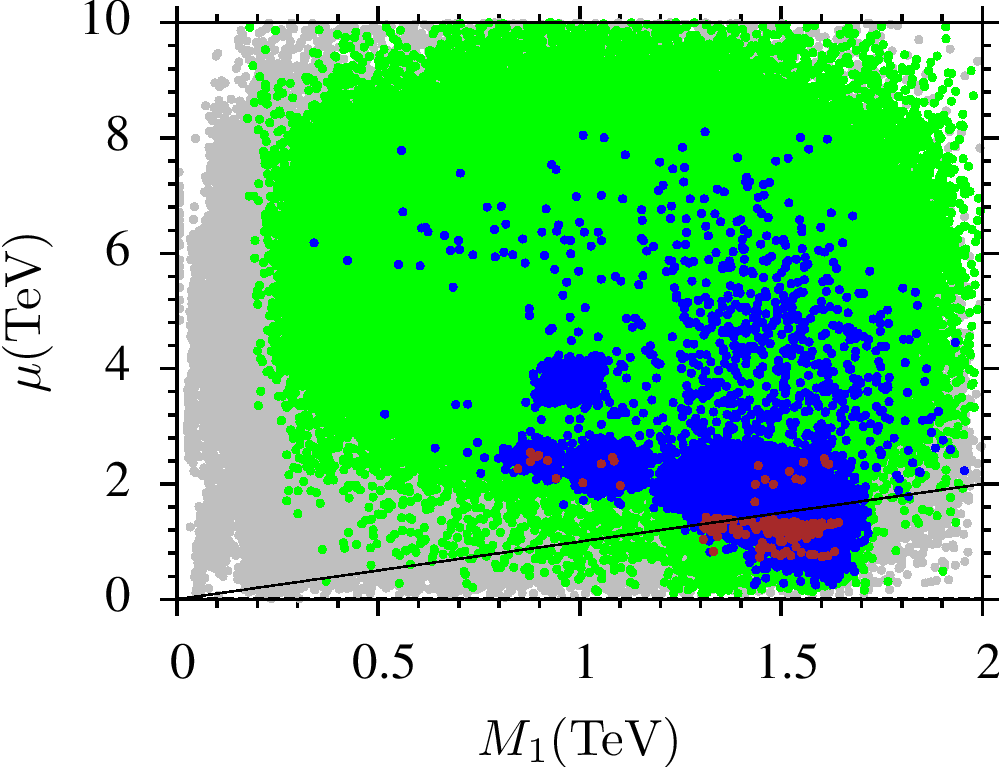} &
\includegraphics[scale=1.2]{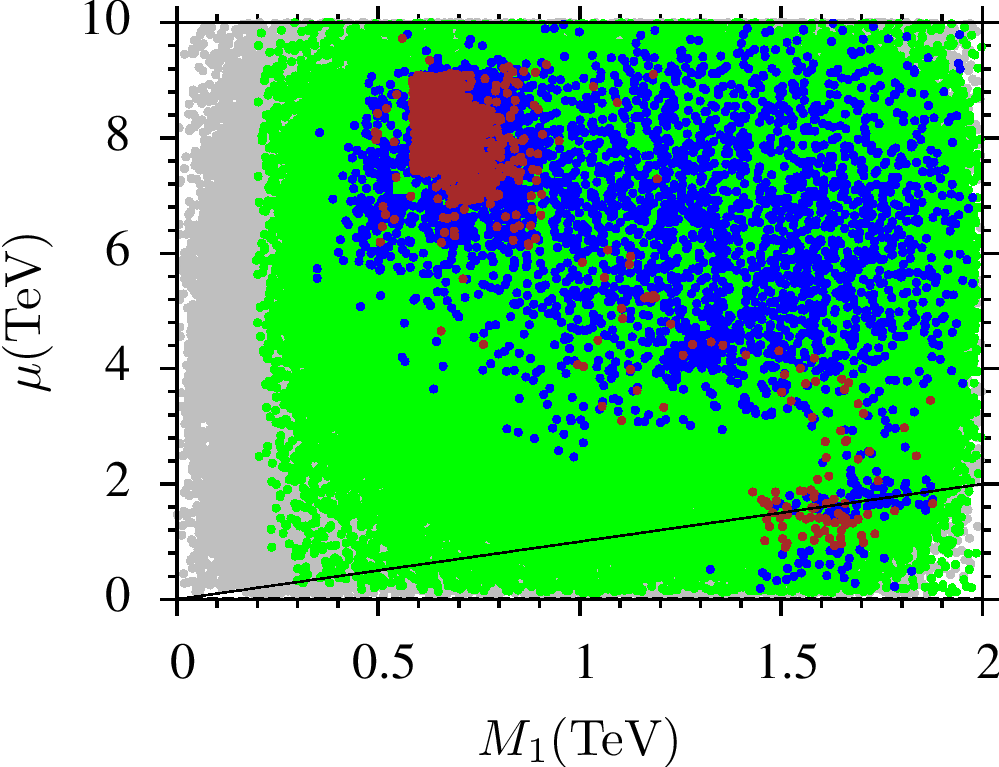}
\end{array}
\end{equation*}
\caption{Plots in $M_{2}-M_{1}$ and $\mu-M_{1}$ planes. Color coding is the same as Fig.\ref{C_spec}.}
\label{DM}
\end{figure}

Fig.\ref{DM} displays the results in $M_{2}-M_{1}$ and $\mu-M_{1}$ planes. Color coding is the same as in Fig.\ref{C_spec}. As stated in the previous section, $M_{2}$ can be as low as 300 GeV. The line in $M_{2}-M_{1}$ plane indicates solutions for which $M_{1}=2M_{2}$ and they yield bino-wino mixing at the low scale. Similarly, the line in $\mu - M_{1}$ plane corresponds to the solutions which have $M_{1} = \mu$. These solutions can lead to very interesting implications, since LSP neutralino is bino-higgsino mixture near this line. Moreover, the LSP neutralino is found to be mostly higgsino below the line. 

\begin{figure}[h!]
\begin{equation*}\hspace{-0.8cm}
\begin{array}{cc}
{\rm ISAJET} & ~~~~~~~~~~~{\rm SoftSusy~ and~ SuperIso Relic} \\ \\
\includegraphics[width=6.4cm,height=4.95cm]{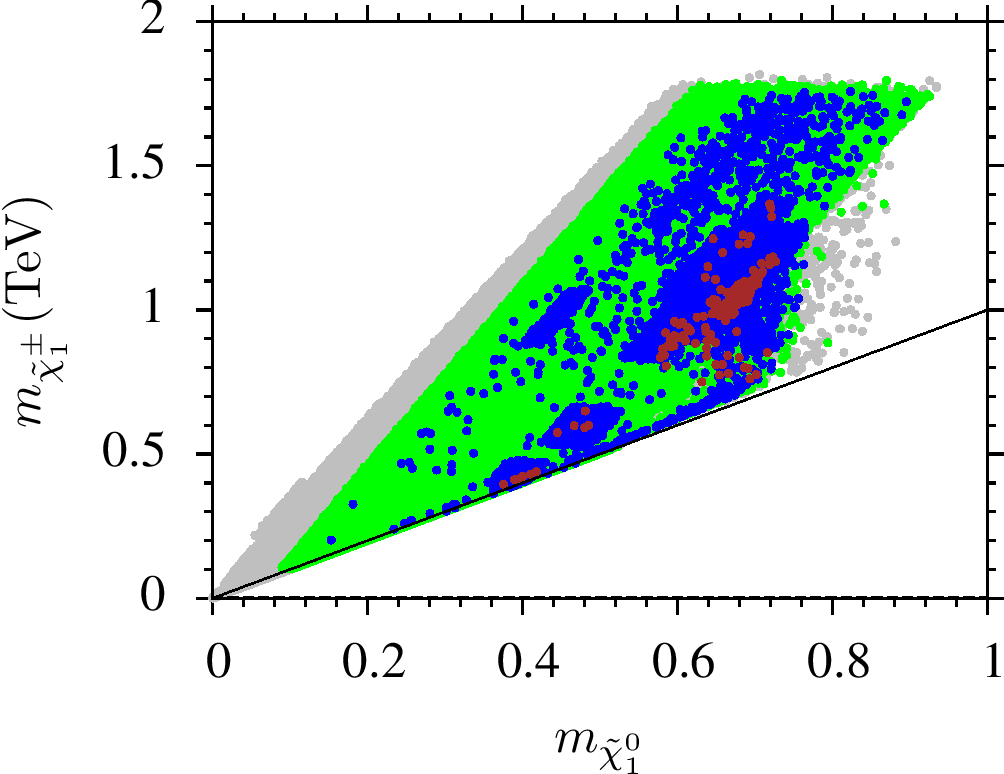} &
\includegraphics[scale=0.9]{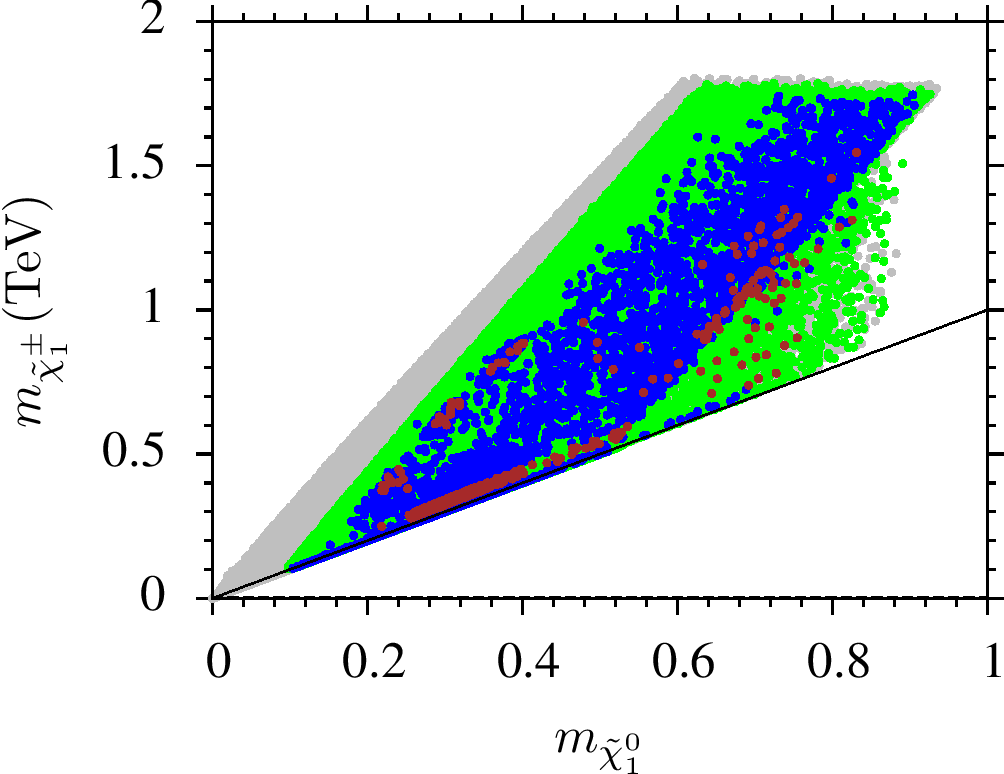} \\
\includegraphics[scale=0.9]{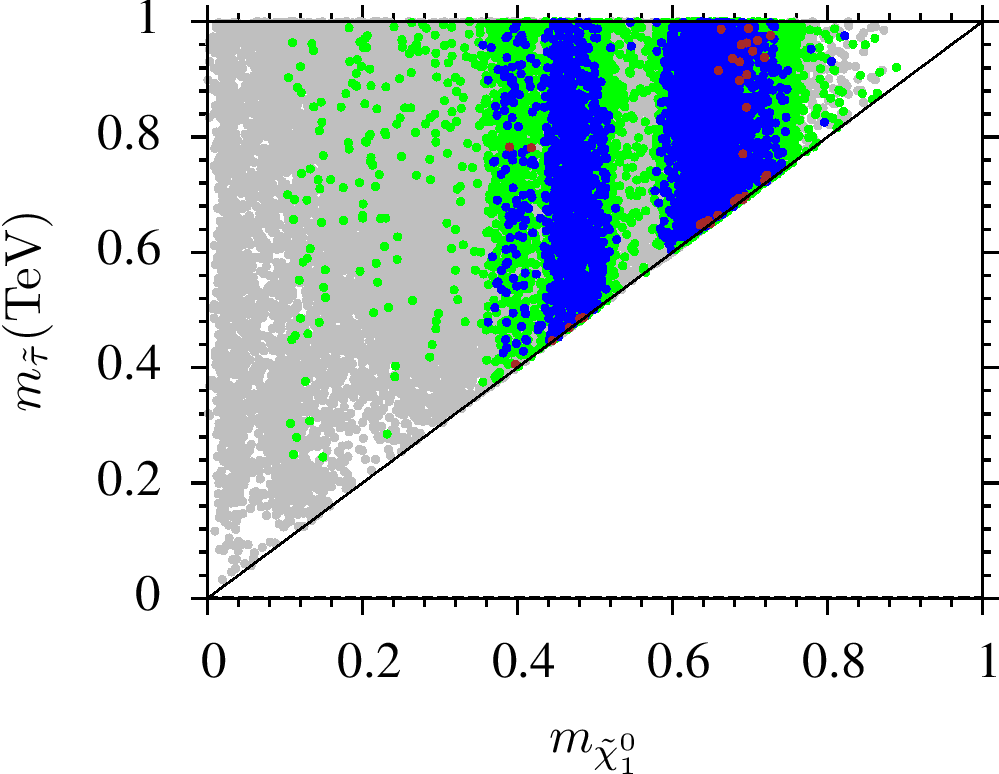} &
\includegraphics[scale=0.9]{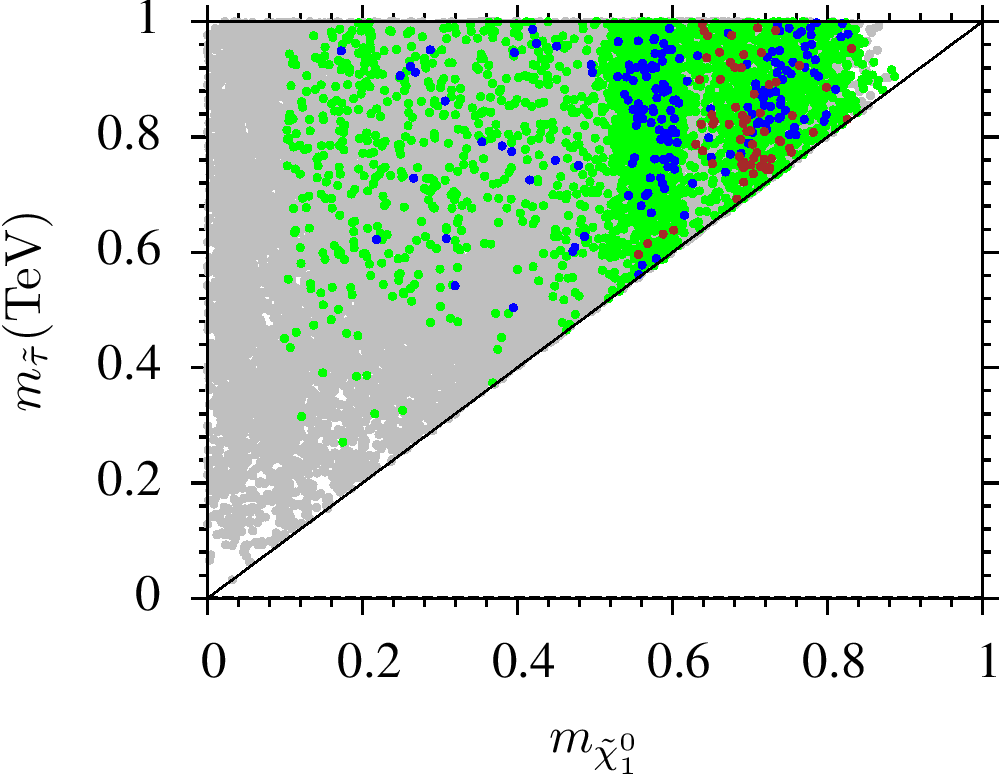} \\
\includegraphics[scale=0.9]{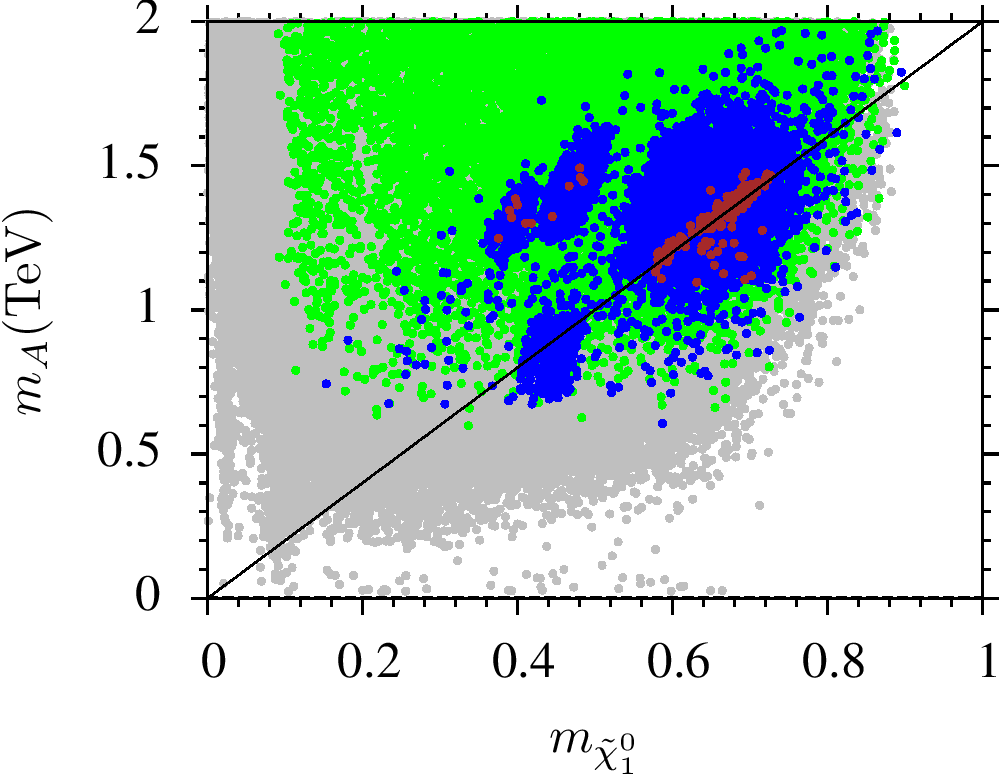} &
\includegraphics[scale=0.9]{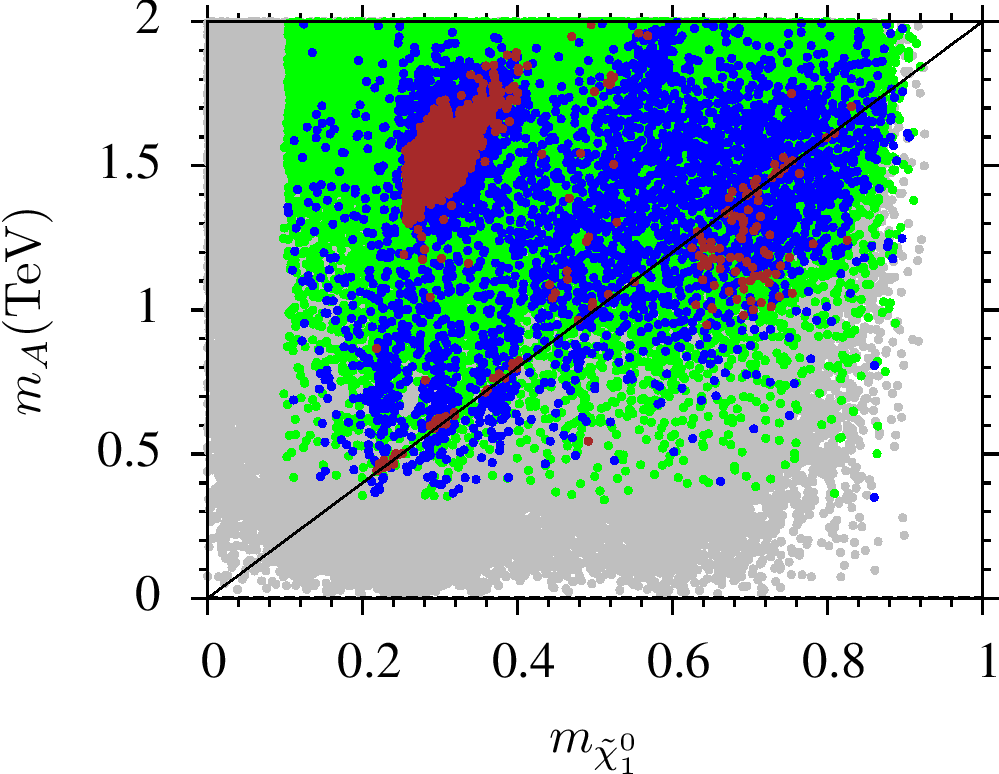}
\end{array}
\end{equation*}
\caption{Plots in $m_{\tilde{\chi}_{1}^{\pm}}-m_{\tilde{\chi}_{1}^{0}}$, $m_{\tilde{\tau}}-m_{\tilde{\chi}_{1}^{0}}$, and $m_{A}-m_{\tilde{\chi}_{1}^{0}}$ planes. Color coding is the same as in Fig.\ref{C_spec}. The solid lines in the plots correspond to the related coannihilation channel regions.}
\label{coan}
\end{figure}

Fig.\ref{coan} summarizes our results for the coannihilation channels compatible with QYU in $m_{\tilde{\chi}_{1}^{\pm}}-m_{\tilde{\chi}_{1}^{0}}$, $m_{\tilde{\tau}}-m_{\tilde{\chi}_{1}^{0}}$, and $m_{A}-m_{\tilde{\chi}_{1}^{0}}$ planes. Color coding is the same as in Fig.\ref{C_spec}. The solid lines in the plots correspond to the related coannihilation channel regions. The ISAJET panel of $m_{\tilde{\chi}_{1}^{\pm}}-m_{\tilde{\chi}_{1}^{0}}$ shows that the neutralino and the lightest chargino of mass $\gtrsim 400$ GeV can be nearly degenerate as expected from the $M_{2}-M_{1}$ planes of Fig.\ref{DM}. We can find solutions with chargino-neutralino coannihilation channel for $m_{\tilde{\chi}_{1}^{\pm}} \simeq m_{\tilde{\chi}_{1}^{0}} \sim 200$ GeV, if we relax the LSP neutralino relic density to $0.0913 \leq \Omega h^{2} \leq 1$, as seen in the SoftSusy and SuperIso Relic panel. Besides chargino-neutralino coannihilation, the stau-neutralino channel is found to be compatible with QYU as seen from the $m_{\tilde{\tau}}-m_{\tilde{\chi}_{1}^{0}}$ plane. There are plenty of solutions for $400 \lesssim m_{\tilde{\tau}}\simeq m_{\tilde{\chi}_{1}^{0}} \lesssim 800$ GeV in the ISAJET panel, while SoftSusy yields fewer solutions with light staus. 

Another solution allowed by QYU is the $A-$resonance shown in the $m_{A}-m_{\tilde{\chi}_{1}^{0}}$ planes. The solid line in these panels corresponds to $m_{A}=2m_{\tilde{\chi}_{1}^{0}}$ in which two LSP neutralinos annihilate via the $A-$boson. The $A-$resonance solutions can be found for  $m_{\tilde{\chi}_{1}^{0}} \gtrsim 600$ GeV in the data set obtained from ISAJET, while it can be realized for $m_{\tilde{\chi}_{1}^{0}} \gtrsim 600$ if one applies $0.0913 \leq \Omega h^{2} \leq 1$ to the data set obtained with SoftSusy and SuperIso.

\section{Higgsino(-like) LSP}
\label{sec:higgsinoDM}

In the previous section we have identified various coannihilation channels and a resonance solution which reduce the relic abundance of LSP neutralino so that the dark matter phenomenology in the 4-2-2 framework can be consistent with the WMAP experiment. In this section we briefly explore an alternative scenario in which the LSP neutralino is a gaugino-higgsino. This case opens up possibilities for direct detection experiments via relic LSP neutralino scattering on nuclei. The case with bino-wino mixture, or equivalently those with chargino-neutralino coannihilation, yield moderate cross-sections in these scattering processes, since the LSP interacts with quarks in the nucleon also via $SU(2)$ interactions. The scattering cross-section reaches its highest values when the LSP neutralino is a bino-higgsino mixture or mostly higgsino, since the Yukawa interactions between quarks and the higgsino component of LSP neutralino take part in the scattering processes. 

\begin{figure}[h!]
\subfigure{\includegraphics[scale=0.6]{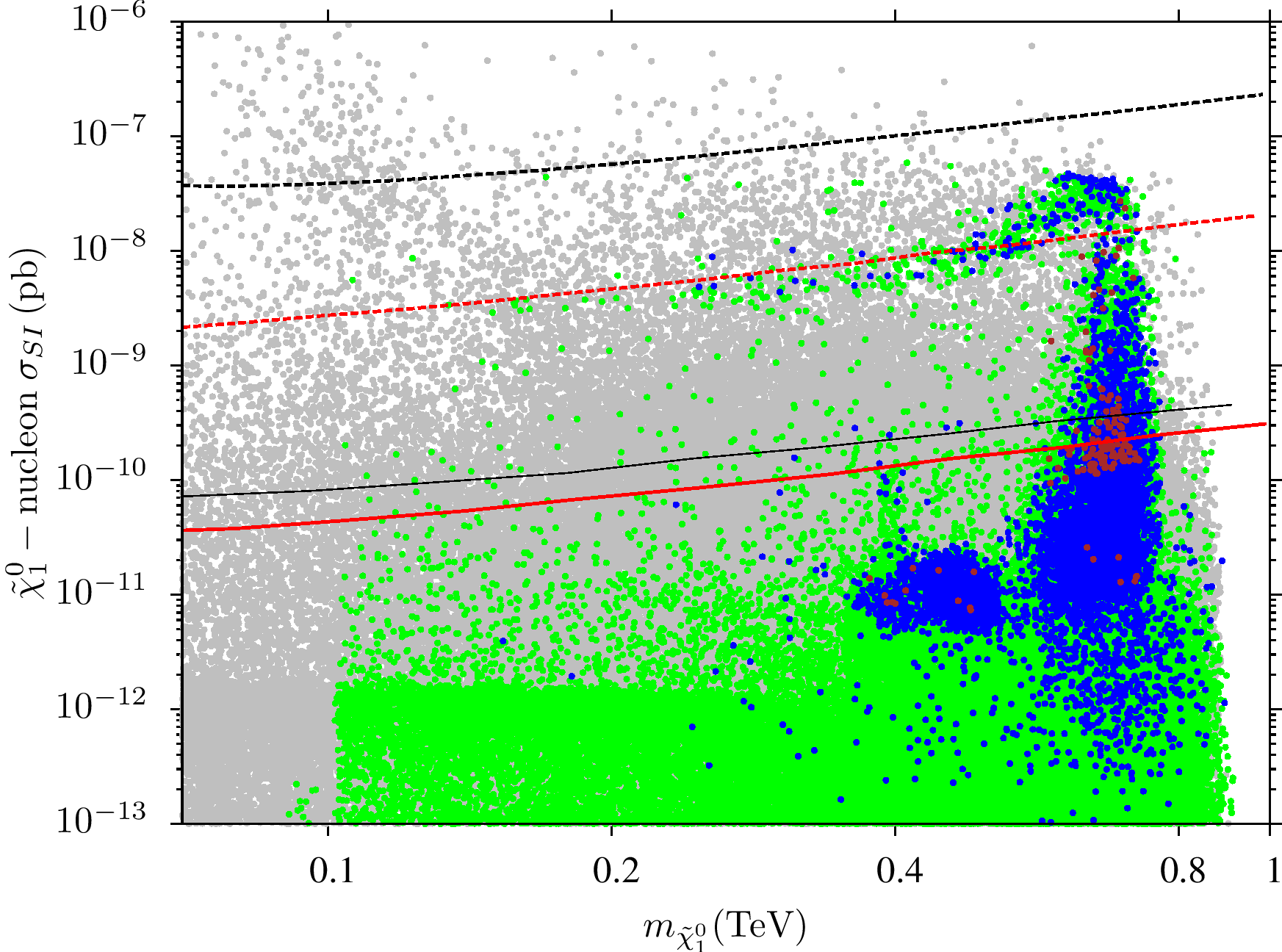}}
\subfigure{\includegraphics[scale=0.6]{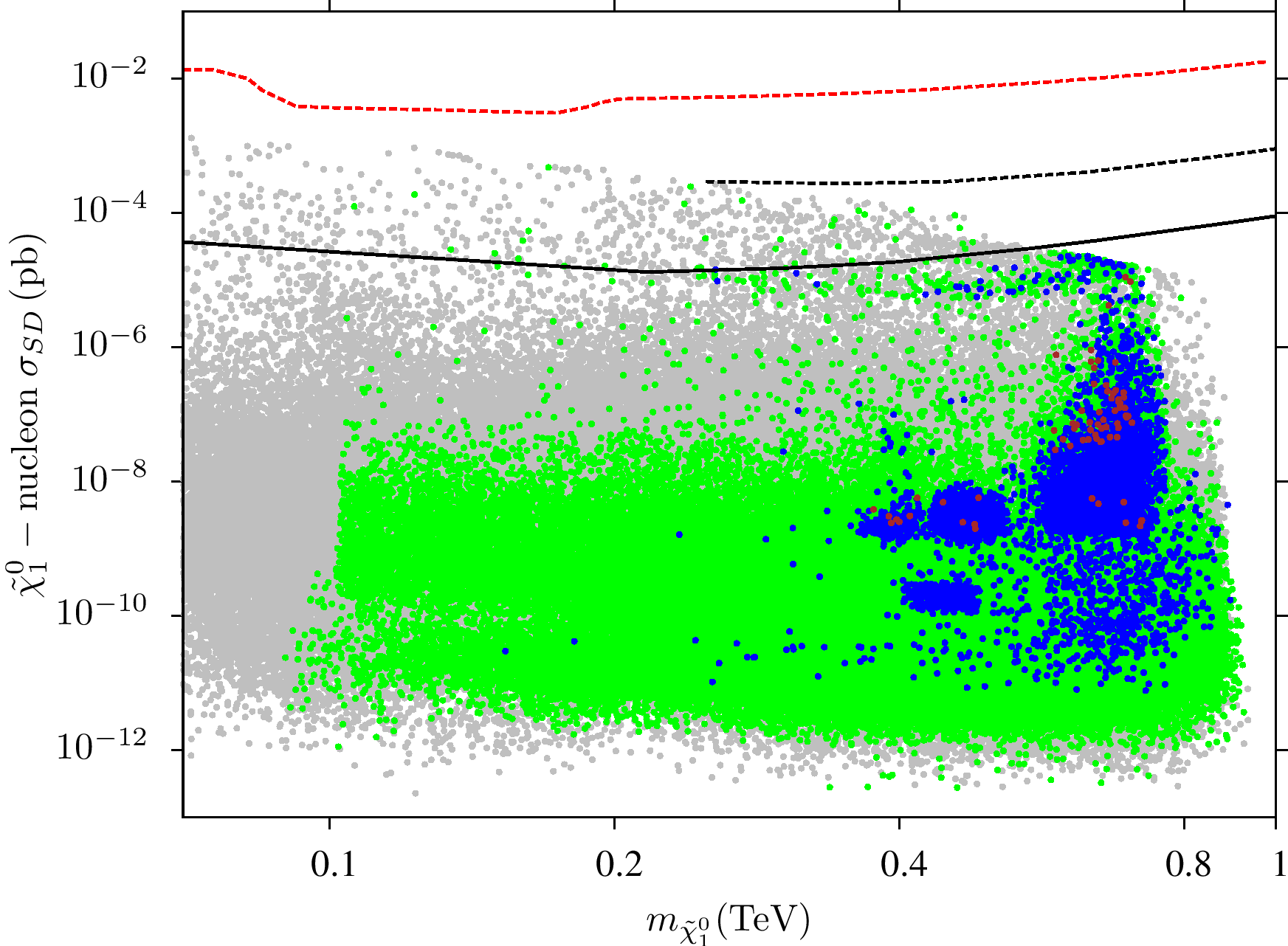}}
\caption{Plots in $\tilde{\chi}_{1}^{0}-{\rm nucleon}~\sigma_{SI}$ and $\tilde{\chi}_{1}^{0}-{\rm nucleon}~\sigma_{SD}$ planes. Color coding is the same as in Fig.\ref{C_spec}. In $\tilde{\chi}_{1}^{0}-{\rm nucleon}~\sigma_{SI}$ plane the dashed (solid) red line represents the current (future) bound of the XENON1T experiment, while the dashed (solid) black line show the current (future) bound of the CDMS experiment. In $\tilde{\chi}_{1}^{0}-{\rm nucleon}~\sigma_{SD}$ plane the dashed red line represents the bound from the Super K experiment, while the dashed (solid) black line shows the current (future) reach of the IceCube experiment.}
\label{directdetection}
\end{figure}

We present our results in Fig.\ref{directdetection} for neutralino-nucleon scattering for both spin-independent and spin-dependent cases in $\tilde{\chi}_{1}^{0}-{\rm nucleon}~\sigma_{SI}$ and $\tilde{\chi}_{1}^{0}-{\rm nucleon}~\sigma_{SD}$ planes. Color coding is the same as in Fig.\ref{C_spec}. In the $\tilde{\chi}_{1}^{0}-{\rm nucleon}~\sigma_{SI}$ plane, the dashed (solid) red line represents the current (future) bound of the XENON1T experiment, while the dashed (solid) black line shows the current (future) bound of the CDMS experiment. In the $\tilde{\chi}_{1}^{0}-{\rm nucleon}~\sigma_{SD}$ plane, the dashed red line represents the bound from the Super K experiment, while the dashed (solid) black line shows the current (future) reach of the IceCube experiment. Only the results from ISAJET are shown in the panels. As seen from the $\tilde{\chi}_{1}^{0}-{\rm nucleon}~\sigma_{SI}$ and $\tilde{\chi}_{1}^{0}-{\rm nucleon}~\sigma_{SD}$ plane, the spin-independent cross-section for the LSP neutralino with bino-wino mixture is of order $10^{-11}$ pb, while it rises by two orders of magnitude for bino-higgsino mixture. Furthermore, the spin-independent cross-section lies between $10^{-10} - 10^{-8}$ pb if the LSP neutralino is mostly a higgsino within reach of the direct detection experiments such as XENON1T and SuperCDMS. Finally, we also check that QYU with Higgsino(-like) dark matter and mass $\sim 1$ TeV is also realized in 4-2-2 and $SO(10)$ supersymmetric models with universal gaugino masses (namely CMSSM boundary conditions) at $M_{{\rm GUT}}$.

\section{Comparison of ISAJET and SoftSusy}
\label{sec:isasoft}

Ref. \cite{Allanach:2003jw} gives a detailed analysis and comparison among several numerical codes including ISAJET and SoftSusy. In Sec.\ref{sec:spec} we show that QYU prefers regions with larger $\tan\beta$, and a $3\%$ difference between the Yukawa couplings obtained from SoftSusy and ISAJET \cite{Allanach:2003jw} can lead to some quantitative differences in the results.

In this section, we present two tables of benchmark points that exemplify the results obtained from our scans. Table \ref{ISAJET} presents four benchmark points obtained from the ISAJET scan. The points are chosen to be consistent with the constraints mentioned in Sec.\ref{sec:scan}. Point 1 is an $A-$resonance solution, and point 2 depicts a solution with higgsino dark matter. Points 3 and 4 display stau-neutralino and chargino-neutralino coannihilation solutions respectively. Point 4 also exemplifies the solution with the heaviest CP-even Higgs boson mass we obtained.

Similarly Table \ref{softsusy} displays four benchmark points consistent with the experimental constraints obtained from SoftSusy and SuperIso Relic scan. The points are chosen to be consistent with the constraints mentioned in Sec.\ref{sec:scan}. Point 1 displays an $A-$resonance solution. Points 2 and 3 depict solutions with the higgsino dark matter, while WMAP bound on relic abundance of LSP neutralino is satisfied through the stau-neutralino coannihilation for Point 2, and chargino-neutralino coannihilation for Point 3. Point 4 also shows a solution with chargino-neutralino coannihilation and exemplifies the solution with the heaviest CP-even Higgs boson mass obtained.

Tables \ref{ISAJET} and \ref{softsusy} also summarize the differences between ISAJET and SoftSusy in the fundamental parameters that yield similar implications. The gaugino masses differ a few hundred GeV in the case of stau-neutralino coannihilation, while it rises up to about 1 TeV in scalar masses $m_{16}$ and $m_{10}$. The largest difference can be realized at the Point 4's which displays the largest SM-like Higgs boson mass from both scan. $m_{16}$ is about 4 TeV heavier in SoftSusy result, while $m_{10}$ is about 2 TeV lighter. On the other hand, the phenomenological results obtained from the scans yield very similar results. The fundamental parameter space allowed by the experimental results and QYU are quite similar, and the same coannihilation channels are identified from both scans. 

\begin{table}[htp!] \hspace{-2.0cm}
\centering
\scalebox{0.85}{
\begin{tabular}{|c|cccc|}
\hline
\hline
&&&&\\
               ISAJET  & Point 1 & Point 2 & Point 3 & Point 4  \\ 
               &&&&\\
\hline
$m_{0}$       & 3362  & 3312 & 2905 & 3844   \\
$M_{1} $      & 1343 & 1615 & 1436 &  893  \\
$M_{2} $      & 1143 & 1407 & 1365 &  480.3   \\
$M_{3} $      & 1643 & 1929 & 1542  & 1512  \\
$m_{10}$      & 4058 & 4377 & 3332 & 4320  \\
$\tan\beta$   & 57.1 & 57.2 & 57.4  & 59.7  \\
$A_0/m_{0}$   & -1.05 & -0.94 & -1.46  &-1.74   \\
$m_t$         & 173.3 & 173.3 & 173.3 & 173.3  \\
$\mu$          & 1420  & {\color{red}752} & 2477 & 1996  \\
\hline
$m_h$            & {\color{red} 123.1}& {\color{red} 123.4}& {\color{red} 123.8} & {\color{red}124.7}  \\
$m_H$           & 1205  & 1126 & 1330  & 1394 \\
$m_A$           & \textbf{1197} & 1118 & 1322  & 1385  \\
$m_{H^{\pm}}$   & 1209 & 1130 & 1334  & 1397 \\

\hline
$m_{\tilde{\chi}^0_{1,2}}$
                 & \textbf{595.8}, 958.5  & \textbf{701}, 766 & \textbf{639}, 1150  & \textbf{ 397.2}, \textbf{413.9}  \\

$m_{\tilde{\chi}^0_{3,4}}$
                 & 959.3, 1343 & \textbf{773},1189 & 2000, 2003  & 2474, 2475 \\

$m_{\tilde{\chi}^{\pm}_{1,2}}$
                & 959.3, 1343 & \textbf{775},1168 & 1151, 2003  & \textbf{414.5}, 2476  \\

$m_{\tilde{g}}$  & 3628 & 4174 & 3399 & 3408  \\
\hline $m_{ \tilde{u}_{L,R}}$
                 & 4533, 4507  & 4860,4816 & 4118, 4061  & 4726, 4737  \\
$m_{\tilde{t}_{1,2}}$
                 & 2772, 3251  & 3044, 3517 & 2388, 2947  & 2395, 3053 \\
\hline $m_{ \tilde{d}_{L,R}}$
                 & 4534, 4501 & 4861, 4816 & 4119, 4054  & 4726, 4737  \\
$m_{\tilde{b}_{1,2}}$
                 & 3223, 3457  & 3489, 3670 & 2915, 3117  & 3028, 3459  \\
\hline
$m_{\tilde{\nu}_{e,\mu}}$
                 &3441 & 3434 & 3036  & 3854 \\
$m_{\tilde{\nu}_{\tau}}$
                 & 2662 &  2647 & 2264  & 2750  \\
\hline
$m_{ \tilde{e}_{L,R}}$
                & 3441, 3395 & 3434, 3362 & 3037, 2951  & 3854, 3856  \\
$m_{\tilde{\tau}_{1,2}}$
                & 1398, 2659 & 1293, 2644 & \textbf{650.7}, 2263 & \textbf{405.6}, 2748  \\
\hline

$\sigma_{SI}({\rm pb})$
                & $ { 0.13\times 10^{-9} } $ & {\color{red} $0.23\times 10^{-7}$} & $0.26\times 10^{-10}  $  & $0.86\times 10^{-11}$  \\

$\sigma_{SD}({\rm pb})$
                &$ { 0.43\times 10^{-7} } $ & $0.94\times 10^{-5}$ & $0.56\times 10^{-8}$  & $0.27\times 10^{-8}$   \\

$\Omega h^{2}$      & 0.108 & 0.104 & 0.128  & 0.106 \\
\hline
&&&&\\
$y_{t,b,\tau}(M_{\rm GUT}) $ & 0.56, 0.41, 0.70 & 0.56, 0.44, 0.70 & 0.55, 0.38, 0.72  & 0.54, 0.37, 0.70 \\
&&&&\\
$ C $  & 0.15 & 0.13 & 0.18  & 0.18 \\
\hline
\hline
\end{tabular}}
\caption{Benchmark points from ISAJET scan. The points are chosen to be consistent with the constraints mentioned in Sec.\ref{sec:scan}. Point 1 is an $A-$resonance solution, Point 2 depicts a solution with the higgsino dark matter, the WMAP bound on relic abundance of LSP neutralino is satisfied through chargino-neutralino coannihilation for this point. Point 3 and Point 4 display stau-neutralino and chargino-neutralino coannihilation solutions respectively.}
\label{ISAJET} 
\end{table}

\begin{table}[t!] \hspace{-2.0cm}
\centering
\scalebox{0.85}{
\begin{tabular}{|c|cccc|}
\hline
\hline
&&&&\\              
 SoftSusy$+$\pbox{20cm}{SuperIso \\ Relic}  & Point 1 & Point 2 & Point 3 & Point 4   \\ 
 &&&&\\
\hline
$m_{0}$       & 1930 &1820  & 2048 &  9832   \\
$M_{1} $      & 1457 &1649 & 1510 &   \textbf{861.2}   \\
$M_{2} $      & 1096 &1470 & 1218 &   \textbf{479.4}   \\
$M_{3} $      & 1999 &1918 & 1949 &  1434  \\
$m_{10}$      & 2873 &2972 & 3189 &  2130  \\
$\tan\beta$   & 54.3 &53.2 & 53.4 &  59.2  \\
$A_0/m_{0}$   &-0.41 & -0.36 & -0.37   &-0.16   \\
$m_t$         & 173.3 & 173.3 & 173.3 & 173.3  \\
$\mu$          & 1678 & {\color{red}962.4}  & {\color{red}903.3} &  7593  \\
\hline
$m_h$            & {\color{red} 123.2} & {\color{red} 123.1}& {\color{red} 123.3} & {\color{red}125}  \\
$m_H$           & \textbf{1259} & 1000  & 948.2  & 1884 \\
$m_A$           & \textbf{1259}  & 1000 & 948.1 &  1884  \\
$m_{H^{\pm}}$   & 1262  &1004 & 952.2 &  1886 \\

\hline
$m_{\tilde{\chi}^0_{1,2}}$
                 & \textbf{635.1}, 897.2  & \textbf{704}, 770  & \textbf{644.1}, 722 &  \textbf{ 386.9}, {\color{red} 413.9}   \\

$m_{\tilde{\chi}^0_{3,4}}$
                 & 1348, 1354 & 778.4, 1219 & 723.8, 1014   & 7569, 7569 \\

$m_{\tilde{\chi}^{\pm}_{1,2}}$
                & 897.3, 1355  & 762, 1219 & \textbf{710.7}, 1014 &  \textbf{413.7}, 7569  \\

$m_{\tilde{g}}$  & 4220 & 4053 & 4129 &  3394  \\
\hline $m_{ \tilde{u}_{L,R}}$
                 & 4086, 4056  & 3963, 3888  & 4082, 4041 &  10040, 10054  \\
$m_{\tilde{t}_{1,2}}$
                 & 3066, 3302  & 2867, 3147  & 2965, 3211 &  7779, 8223 \\
\hline $m_{ \tilde{d}_{L,R}}$
                 & 4087, 4048  & 3964, 3876 & 4084, 4033 &  10041, 10056  \\
$m_{\tilde{b}_{1,2}}$
                 & 3272, 3377  & 3117, 3154  & 3185, 3259 &  8215, 8661  \\
\hline
$m_{\tilde{\nu}_{e,\mu}}$
                 & 2054, 2053  & 2055, 2053 & 2192, 2191 &  9817, 9813 \\
$m_{\tilde{\nu}_{\tau}}$
                 & 1621 & 1632  & 1734 &  8719  \\
\hline
$m_{ \tilde{e}_{L,R}}$
                & 2056, 2003  & 2056, 1918 & 2194, 2121 &  9817, 9829  \\
$m_{\tilde{\tau}_{1,2}}$
                & 899.1, 1627 & \textbf{735.7}, 1635 & 937.3, 1737 &  7489, 8722  \\
\hline
&&&&\\
$\Omega h^{2}$      & 0.103  & 0.13 & 0.095 &  0.092 \\
&&&&\\
\hline
&&&&\\
$y_{t,b,\tau}(M_{\rm GUT}) $ & 0.56, 0.44, 0.67  & 0.56, 0.46, 0.66 & 0.56, 0.46, 0.65 & 0.54, 0.40, 0.67 \\
&&&& \\
$ C $  & 0.12  & 0.10 & 0.09 &  0.15 \\
\hline
\hline
\end{tabular}}
\caption{Benchmark points from SoftSusy and SuperIso Relic scan. The points are chosen to be consistent with the constraints mentioned in Sec.\ref{sec:scan}. Point 1 displays an $A-$resonance solution. Points 2 and 3 depict solutions with the higgsino dark matter, while WMAP bound on relic abundance of LSP neutralino is satisfied through the stau-neutralino coannihilation for Point 2, and chargino-neutralino coannihilation for Point 3. Point 4 also shows a solution with chargino-neutralino coannihilation and exemplifies the solution with the heaviest CP-even Higgs boson mass obtained.}
\label{softsusy} 
\end{table}

\section{Conclusion}
\label{sec:conclusion}

We have employed ISAJET and SoftSusy interfaced with SuperIso relic to explore the LHC implications of Quasi-Yukawa unified (QYU) supersymmetric models based on $G=SU(4)_{c}\times SU(2)_{L}\times SU(2)_{R}$. In these QYU models, the  third family Yukawa unification relations involving $t$, $b$ and $\tau$, is quantified by a parameter $C$ which takes values $\sim 0.1 - 0.2$. In contrast to earlier studies, the MSSM gaugino masses at $M_{{\rm GUT}}$ are non-universal but consistent with the gauge symmetry $G$. The thermal relic abundance of the LSP neutralino is compatible with the WMAP bounds through the chargino and stau coannihilation channels, as well as the A-resonance solution. We also identify solutions with Higgsino-like and pure Higgsino dark matter ($\mu \lesssim 1$ TeV) which can be tested in the direct dark matter searches such as XENON1T and SuperCDMS. The predicted gluino mass ranges from 1-4 TeV, while the stop masses are heavier than 2 TeV or so. It is reassuring to note that the low energy phenomenology obtained from ISAJET and SoftSusy are in good qualitative agreement.

\section*{Acknowledgment}

We thank Ilia Gogoladze, M. Adeel Ajaib and Dimitri I. Kazakov for useful discussions. \c{S}HT would like to thank Michael Afanasev for his advice in programming. This work is supported in part by DOE Grants DE-FG02-91ER40626 (QS) and Russian Foundation for Basic Research (RFBR) Grant No. 14-02-00494-a (\c{S}HT). This work used the Extreme Science and Engineering Discovery Environment (XSEDE), which is supported by the National Science Foundation grant number OCI-1053575. Part of the numerical calculations reported in this paper were performed at the National Academic Network and Information Center (ULAKBIM) of Turkey Scientific and Technological Research Institution (TUBITAK), High Performance and Grid Computing Center (TRUBA Resources).

\end{document}